\documentclass[traditabstract]{aa}
\bibpunct{(}{)}{;}{a}{}{,} 
\usepackage{graphicx}
\usepackage[varg]{txfonts}
\usepackage{amsmath}
\usepackage{color}

\begin{document}

\title{Constraining the geometry and kinematics of the quasar broad emission line region using gravitational microlensing.\\I. Models and simulations}
\author{L. Braibant\inst{1,}\thanks{Research Assistant  F.R.S.-FNRS},
        D. Hutsem\'ekers\inst{1,}\thanks{Senior Research Associate F.R.S.-FNRS}, 
        D. Sluse\inst{1},
        R. Goosmann\inst{2}
        }
\institute{
    Institut d'Astrophysique et de G\'eophysique,
    Universit\'e de Li\`ege, All\'ee du 6 Ao\^ut 19c, B5c,
    4000 Li\`ege, Belgium
    \and 
    Observatoire Astronomique de Strasbourg,
    Universit\'e de Strasbourg, Rue de l’Universit\'e 11,
    F-67000 Strasbourg, France
    }
\date{Received ; accepted: }
\titlerunning{Microlensing of the quasar broad emission line region. I. Models} 
\authorrunning{L. Braibant et al.}
\abstract{
  Recent studies have shown that line profile distortions are commonly observed in gravitationally lensed quasar spectra. Often attributed to microlensing differential magnification, line profile distortions can provide information on the geometry and kinematics of the broad emission line region (BLR) in quasars.  We investigate the effect of gravitational microlensing on quasar broad emission line profiles and their underlying continuum, combining the emission from simple representative BLR models with generic microlensing magnification maps. Specifically, we considered Keplerian disk, polar, and equatorial wind BLR models of various sizes.  The effect of microlensing has been quantified with four observables: $\mu^{BLR}$, the total magnification of the broad emission line; $\mu^{cont}$, the magnification of the underlying continuum; as well as red/blue, $RBI$ and wings/core, $WCI$, indices that characterize the line profile distortions. The simulations showed that distortions of line profiles, such as those recently observed in lensed quasars, can indeed be reproduced and attributed to the differential effect of microlensing on spatially separated regions of the BLR. While the magnification of the emission line $\mu^{BLR}$ sets an upper limit on the BLR size and, similarly, the magnification of the continuum $\mu^{cont}$  sets an upper limit on the size of the continuum source, the line profile distortions mainly depend on the BLR geometry and kinematics.  We thus built $(WCI,RBI)$ diagrams that can serve as diagnostic diagrams to discriminate between the various BLR models on the basis of quantitative measurements. It appears that a strong microlensing effect puts important constraints on the size of the BLR and on its distance to the high-magnification caustic. In that case, BLR models with different geometries and kinematics are more prone to produce distinctive line profile distortions for a limited number of caustic configurations, which facilitates their discrimination. When the microlensing effect is weak, there is a larger overlap between the characteristics of the line profile distortions produced by the different models, and constraints can only be derived on a statistical basis.
}
\maketitle
%
%
%
\section{Introduction}
\label{sec:intro}

The broad emission lines (BELs) that dominate quasar optical spectra constitute one of their main observational characteristics. Since the region at the origin of these lines, the broad emission line region (BLR), is unresolved with current instrumentation, its geometry and kinematics can only be investigated through indirect methods. Reverberation mapping \citep{1982BlandfordMcKee,1993Peterson} takes advantage of the variability of the ionizing continuum to measure the response time of the emission lines, which is directly related to the size of the BLR. One of the most important results of reverberation mapping is a precise measurement of the BLR radius - AGN luminosity relationship, as expected from simple photoionization arguments \citep{2005Kaspi,2010Bentz}. With recent high quality data, time lags as of function of the velocity across the H$\beta$ emission line profile have been measured in about 10 AGNs providing information on the geometry and kinematics of the low-ionization BLR. Various kinematical signatures have been found in the different objects; these kinematical signatures are mostly virialized motions and inflows but also outflows \citep{2009Bentz, 2010Bentz, 2010Denney, 2014Pancoast, 2016Du, 2017Grier}.

By selectively magnifying different subregions of the BLR, gravitational microlensing can also provide information on the size, geometry, and kinematics of the BLR in quasars \citep{1988Nemiroff, 1990SchneiderWambsganss}. Microlensing-induced line profile deformations are now commonly observed in gravitationally lensed quasar spectra, exhibiting various symmetric and asymmetric distortions in both low- and high-ionization line profiles \citep{2004Richards, 2005Wayth, 2007Sluse, 2011Sluse, 2012Sluse, 2011ODowd, 2013Guerras, 2014Braibant, 2016Braibant}. The size of the BLR has been estimated in a few objects and found compatible with reverberation mapping measurements \citep{2005Wayth, 2011Sluse, 2013Guerras}. Microlensing can thus be a powerful and alternative approach to reverberation mapping, especially since it can be applied to high redshift quasars, with little dependence on their luminosity, and to the study of both the low- and high-ionization BLRs.

Making use of a dedicated line profile disentangling technique, \cite{2014Braibant, 2016Braibant} have extracted the part of the emission line profile affected by microlensing in two lensed quasars for which high quality data were available. A strong difference in the distortions suffered by the \ion{C}{iv} and H$\alpha$ BELs was uncovered, indicating that high- and low-ionization emission lines must originate from different regions. From the detection of a red/blue differential microlensing effect in the H$\alpha$ line profile these authors inferred that the low-ionization BLR is likely a Keplerian disk, while the wings/core distortion observed in the \ion{C}{iv} line can be interpreted assuming a polar wind high-ionization BLR. However, the mapping between a wavelength range in the line profile and subregions in the BLR seen in projection is usually not unique so that a confrontation of observations to detailed modeling appears necessary.  Reconstructing the BLR from the microlensing signal is a complex task since the line profile distortions induced by microlensing depend not only on the BLR velocity field and geometry but also on the microlensing magnification pattern, which may be an intricate caustic structure not directly observable.

Possible effects of microlensing on BELs have already been theoretically investigated by several authors \citep{1988Nemiroff, 1990SchneiderWambsganss, 1994Hutsemekers, 2001Popovic, 2002Abajas, 2007Abajas, 2004LewisIbata, 2011ODowd, 2014Simic} considering various BLR models and magnification patterns \citep[see][for a review]{2012Sluse}. In this paper, we extend those works, focusing on the interpretation of the observed line profile distortions caused by microlensing, such as those reported in \cite{2014Braibant, 2016Braibant}.  The main purpose of our study is to determine whether simple BLR and microlensing models can reproduce the observations and then if it is possible to discriminate between the BLR models based on the observed microlensing-induced line profile distortions. Special emphasis is given to the ability of microlensing to produce either symmetric or asymmetric distortions in the line profiles.

Modeling the effect of gravitational microlensing on broad emission line profiles is a two-step process that involves, first, simulating the emission of the BLR at different wavelengths for a given line of sight (Sect.~\ref{sec:modelingBLR}) and, then, convolving each monochromatic snapshot of the BLR by caustic structures that represent the spatially inhomogeneous magnification caused by stars in the lens galaxy (Sect.~\ref{sec:modelingMicrolensing}). Since microlensing caustics simultaneously magnify the BLR and the source of continuum, models of the continuum source are presented in Sect.~\ref{sec:continuum_source}. To simplify the analysis and interpretation of the simulations and allow comparison with observations, observables characterizing line profile distortions are defined in Sect.~\ref{sec:measure_distortions}. The results of the simulations and general trends are discussed in Sect.~\ref{sec:modelingBLR-results}. We describe our conclusions in the last section. In subsequent papers,  models and simulations will be confronted to the observations reported in \cite{2014Braibant, 2016Braibant} with the goal of constraining the geometry and kinematics of the BLR.

\section{Modeling the BLR}
\label{sec:modelingBLR}
\subsection{\texttt{STOKES}}
\label{sec:stokes}

We used the Monte Carlo radiative transfer code \texttt{STOKES} \citep{2007Goosmann, 2012Marin, 2014Goosmann} to model the emission of the BLR at different Doppler velocity shifts. \texttt{STOKES} is a flexible code that allows one to define an emission region with a parameterizable cylindrical or conical geometry, and a three-dimensional velocity field. \texttt{STOKES} implicitly assumes that the emission region is optically thin and that the line broadening is only due to the Doppler velocity of the BLR gas. In the following, we describe the adopted parameters and modifications made to the code to build representative BLR models.

\subsubsection{Geometry of the BLR}

\begin{figure}
\centering
\resizebox{\hsize}{!}{\includegraphics*{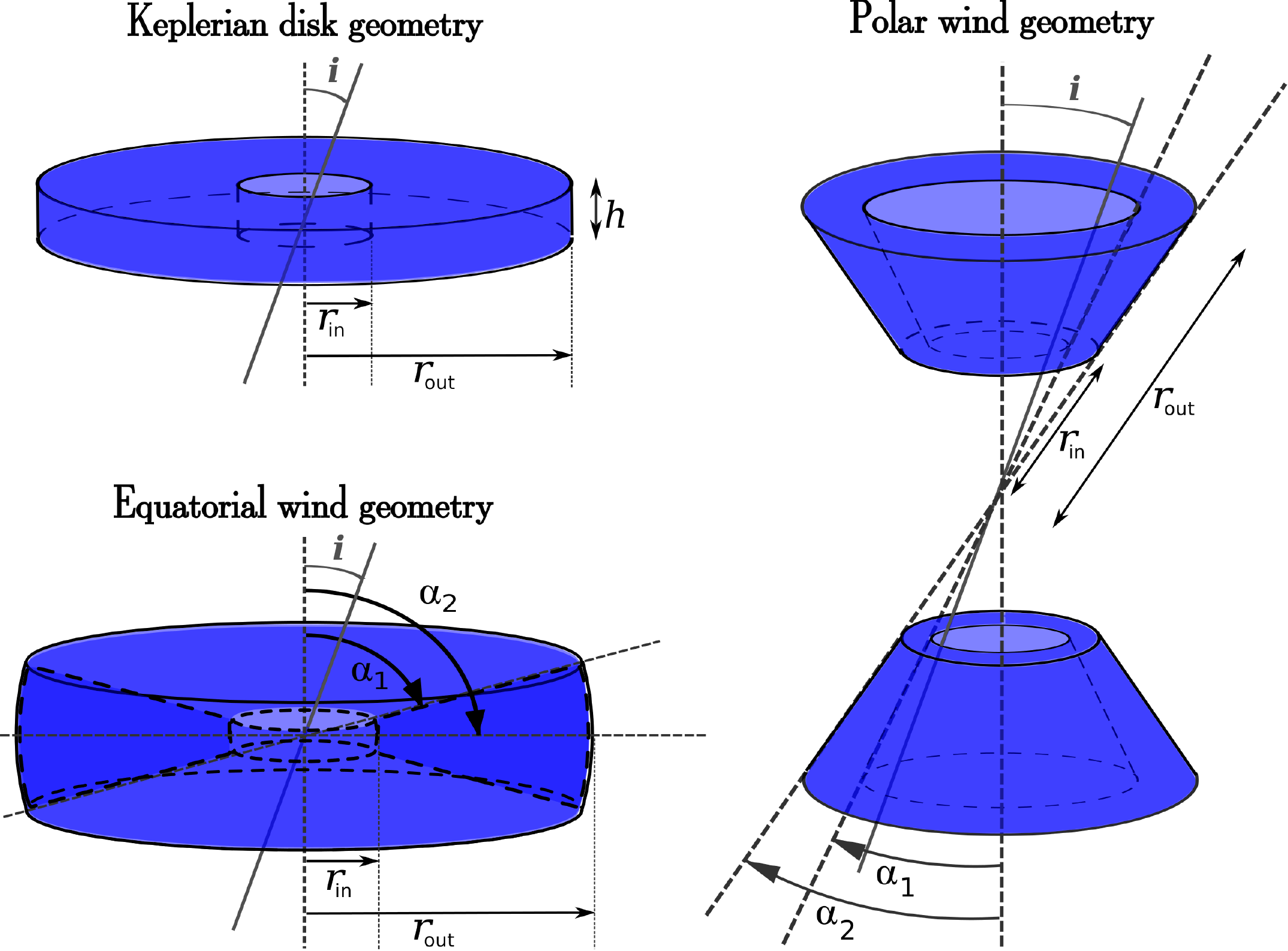}}
\caption{Geometry of the BLR models investigated. \texttt{STOKES} allows us to parameterize cylindrical and conical emission regions. The thin Keplerian disk geometry is obtained by taking a height that is much smaller than the outer radius $r_{\text{out}}$ (\textit{upper left}). The polar wind geometry is a thin hollow bicone with an opening angle going from $\alpha_1 = 45\degr$ to $\alpha_2 = 60\degr$ (\textit{right}). The equatorial wind geometry is built using the same conical geometry as for the polar wind but with $\alpha_1 = 75\degr$ and $\alpha_2 = 90\degr$, resulting in a flared disk with an opening angle of $30 \degr$ (\textit{lower left}).}
\label{fig:models_geometry}
\end{figure}

Recent microlensing studies have unveiled differential microlensing magnification of the red and blue parts of BELs in several systems \citep{2004Richards, 2012Sluse, 2014Braibant, 2016Braibant}. We therefore discard spherically symmetric models from our analysis since negative and positive velocities superpose on the line of sight, preventing any asymmetric red/blue magnification.

To model the BLR, we consider hereafter three simple though representative geometries and kinematics commonly invoked in literature \citep[e.g.,][]{1995Robinson, 2002Abajas}: a polar radially expanding wind (PW), an equatorial radially expanding wind (EW), and a Keplerian rotating disk (KD). The geometry of the models is illustrated in Fig.~\ref{fig:models_geometry}. In line with previous studies, we assume that the number of emitting clouds is large enough to consider the gas homogeneous. We hypothesize that the conditions of temperature and ionization that enable line emission are reached at some distance from the central ionizing continuum source, which is mimicked by an ``inner BLR radius'', $r_{\text{in}}$. This confines the line-emitting gas to a shell centered on the continuum source. The inner radius of the BLR is used as the reference scale in the definition of the spatial and velocity structures.

For each model, \texttt{STOKES} provides the surface brightness of the BLR in different velocity intervals (i.e., monochromatic maps), and for several inclinations --polar viewing angles-- of the line of sight with respect to the symmetry axis of the system ($i=0 \degr$ when the axis is parallel to the line of sight).

We emphasize that these models highlight key features of the BLR that could be constrained by microlensing, making use of a minimum number of free parameters. A combination of these simple models and ultimately more physically realistic models might be needed to interpret the large data sets expected from long-term high-frequency spectrophotometric monitoring of lensed quasars.

\subsubsection{Velocity field of the BLR}

The original parameterization of the radial, $v_r$, and azimuthal velocity, $v_{\phi}$, in the \texttt{STOKES} code is modified such that these velocities can vary as a function of the radial distance, $r$. The new parameterization is
\begin{equation}
v_r (r) = v_{\infty} \left( 1 - r_f / r \right)^{\beta} \hspace{5mm}
\label{eq:vradial}
\end{equation}
for the wind models (PW, EW), and
\begin{equation}
v_{\phi} (r, i) = v_{\phi,0} \sin(i) \sqrt{\frac{r_{\text{in}}}{r}} \hspace{5mm}
\label{eq:vazim}
\end{equation}
for the rotating disk model (KD).  The velocity law given in Eq.~\eqref{eq:vradial} is taken from \citet{1995Murray}. It represents the velocity of a wind driven out of the accretion disk with the approximation of simple radial dependence. Large radial acceleration occurs at small radii. The velocity then increases smoothly to reach the velocity at large distance, $v_{\infty} $. The parameter $\beta$ is fixed to 0.5 and the wind launching radius to $r_f = 0.95 r_{\text{in}}$. The azimuthal velocity, given in Eq.~\eqref{eq:vazim}, evolves with both the radial distance and polar angle, $i$, as the pseudo-Keplerian motion defined in \citet{2012Goad}. The maximum velocity, $v_{\phi,0}$, is reached at the inner radius. The variation of the radial and azimuthal velocities with the radial distance is shown in Fig.~\ref{fig:velo+emissivity}.

\subsubsection{Emissivity of the BLR}
\label{sec:emissivity}

\begin{figure}
\centering
\resizebox{\hsize}{!}{\includegraphics*{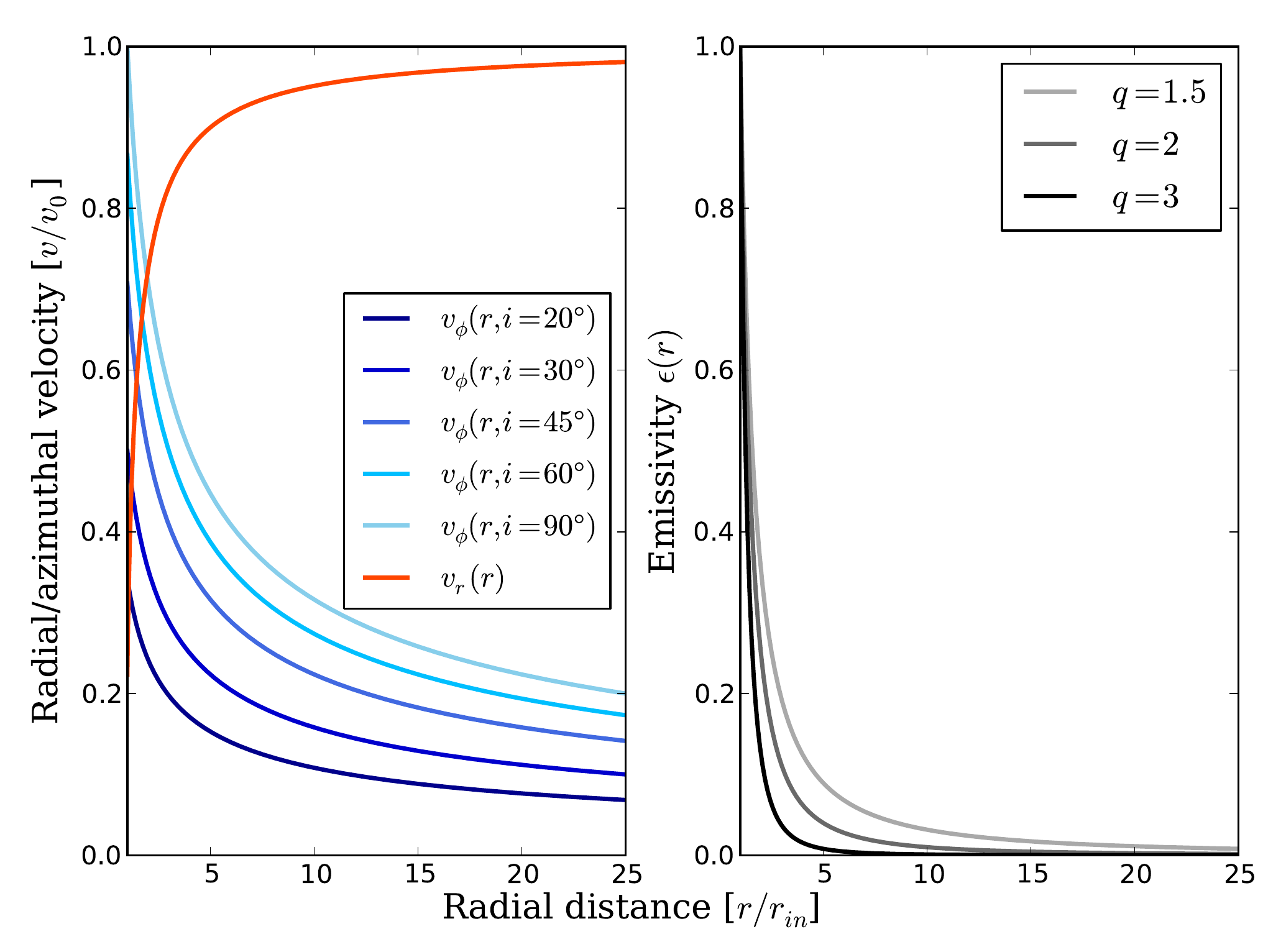}}
\caption{\textit{Left:} Evolution of the radial velocity, $v_r$, and azimuthal velocity, $v_{\phi}$, with the radial distance, $r$. \textit{Right:} Variation of the BLR emissivity as $\epsilon (r) = \epsilon_0 \, (r/r_{\text{in}})^{-q}$ for different values of the power-law exponent $q$.}
\label{fig:velo+emissivity}
\end{figure}

The \texttt{STOKES} code starts the simulation of a photon journey by randomly sampling its initial position within the volume of the BLR. The Monte Carlo method is used to generate a random event $x$ that is characterized by a probability density distribution $p(x)$ defined in the interval $[0,x_{max}]$. The sampling equation relates a number $t$ uniformly and randomly sampled in the $[0,1]$ interval to $x$ through its cumulative distribution function (CDF) $P(x) = \int_0^x p(x') \, \text{d}{ x'}$, that is also defined over $[0,1]$ as follows:
\begin{equation}
t = P(x) = \frac{\int_0^x p(x') \, \text{d}x'}{\int_0^{x_{max}} p(x') \, \text{d}x'} \hspace{3mm} .
\label{eq:sampling_equation}
\end{equation} 
The value of $x$ as a function of $t$ is obtained by inverting this equation.

In its basic version, \texttt{STOKES} assumes a constant density of the emitting material. Following \citet{1995Robinson} and \citet{2002Abajas}, this is modified to account for the variation of the emissivity as a power law of the radial distance, $\epsilon (r) = \epsilon_0 \, (r/r_{\text{in}})^{-q}$. Since the emissivity is assumed to only depend on the radial distance, this only impacts the sampling equation for the initial radial coordinate of the emitted photon.

In the cylindrical symmetry that characterizes the Keplerian disk model, the CDF of the emission radius, $r_{\text{em}}$, of a photon is
\begin{equation}
P(r_{\text{em}}) = \frac{\int_{0}^{2 \pi} \int_{-h/2}^{h/2} \int_{r_{\text{in}}}^{r_{\text{em}}} \epsilon_0 \, (r/r_{\text{in}})^{-q} \, r \, \, \text{d}r \, \text{d}z \, \text{d}\phi}{\int_{0}^{2 \pi} \int_{-h/2}^{h/2} \int_{r_{\text{in}}}^{r_{\text{out}}} \epsilon_0 \, (r/r_{\text{in}})^{-q} \, r \, \, \text{d}r \, \text{d}z \, \text{d}\phi} \hspace{2mm},
\end{equation}
where $(r,\phi,z)$ are the radius, azimuthal angle, and height that define the position in a cylindrical coordinate system. This yields to a sampling equation for the radial distance of the emitted photon given by
\begin{align}
r_{\text{em}} = & \hspace{1mm} \left( t \, r_{\text{out}}^{2-q} + (1-t) \, r_{\text{in}}^{2-q} \right)^{1/(2-q)}  & \text{for} \hspace{1mm}q \neq 2 , \nonumber \\
r_{\text{em}} = & \hspace{1mm} r_{\text{in}} \left( r_{\text{out}} / r_{\text{in}} \right)^t  & \text{for} \hspace{1mm} q = 2 ,
\label{eq:sampling_r_cylinder}
\end{align}
where $r_{\text{in}}$ and $r_{\text{out}}$ are the inner and outer radii of the BLR, respectively  (Fig.~\ref{fig:models_geometry}).
For the radially expanding wind models, which are naturally described in a spherical coordinate system $(r,\phi,i)$, the CDF of the emission radius is
\begin{equation}
P(r_{\text{em}}) = \frac{\int_{0}^{2 \pi} \int_{\alpha_1}^{\alpha_2} \int_{r_{\text{in}}}^{r_{\text{em}}} \epsilon_0 \, (r/r_{\text{in}})^{-q} \, r^2 \, \sin i \, \, \text{d}r \, \text{d}i \, \text{d}\phi}{\int_{0}^{2 \pi} \int_{\alpha_1}^{\alpha_2} \int_{r_{\text{in}}}^{r_{\text{out}}} \epsilon_0 \, (r/r_{\text{in}})^{-q} \, r^2 \, \sin i \, \text{d}r \, \, \text{d}i \, \text{d}\phi} \hspace{2mm},
\end{equation}
such that the sampling equation of the radial distance of the emitted photon is
\begin{align}
r_{\text{em}}  = & \hspace{1mm} \left( t \hspace{1mm}r_{\text{out}}^{3-q} + (1-t) r_{\text{in}}^{3-q} \right)^{1/(3-q)}  & \text{for} \hspace{1mm}q \neq 3 , \nonumber \\
r_{\text{em}}  = & \hspace{1mm} r_{\text{in}}  \left( r_{\text{out}} / r_{\text{in}} \right)^t  & \text{for} \hspace{1mm} q = 3 .
\label{eq:sampling_r_cone}
\end{align}

Photoionization calculations performed for different BELs by \citet{2012Goad} suggest $1.5<q<3$. In their modeling, \citet{2002Abajas} used $q=1.5$. Disk illumination by a central source corresponds to $q=3$ and is widely used in the literature \citep{1994Eracleous, 2009Bon}. We therefore consider $q$ as a free parameter with values $q = 1.5$ and $q = 3$. To emphasize the effect of microlensing, we consider hereafter the compact emissivity law ($q = 3$), and discuss the effect of using $q=1.5$ in Sect. \ref{sec:impact_q}. Figure~\ref{fig:velo+emissivity} shows that, when the emissivity decreases with the radius as $r^{-3}$, most of the emission occurs within $5 \, r_{\text{in}}$ and becomes negligible at a distance of $10 \, r_{\text{in}}$. The emission of the BLR is thus simulated between $r_{\text{in}}$ and $r_{\text{out}} = 10 \,r_{\text{in}}$. Maps of monochromatic BLR emission are computed over a $20 \, r_{\text{in}} \times 20 \, r_{\text{in}}$ area.

\subsubsection{Spatial and spectral samplings}
\label{sec:spaspesam}

The emission from the BLR is collected by a web of virtual detectors that are distributed on a spherical surface around the central source. The detectors are uniformly distributed along the azimuthal coordinate and are distributed according to a cosine distribution along the polar coordinate. Each detector captures the photons whose propagation direction is included in the solid angle it samples. The emission position is then projected onto a distant sky plane, which is orthogonal to the line of sight toward the detector. That grid of virtual observers must be sufficiently fine to limit the image distortion. We consequently pick up 40 azimuthal and 40 polar viewing directions.

Each detector collects BLR photons over a $20 \, r_{\text{in}} \times 20 \, r_{\text{in}}$ spatial area, sampled by $500 \times 500$ pixels, and over the $[-v_0,v_0]$ velocity (spectral) range, where $v_0$ is the highest velocity that can be reached, i.e., $v_{\phi,0}$ for a pure Keplerian motion (Eq.~\ref{eq:vazim}) and $v_{\infty}$ for a pure radial outflow (Eq.~\ref{eq:vradial}). That spectral range is divided into $20$ velocity slices. This ensures that the emission line profiles are sampled by at least six spectral bins, especially the narrower line profiles that originate from a Keplerian disk or from an equatorial wind seen nearly face-on. 

\begin{figure*}
\centering
\resizebox{\hsize}{!}{\includegraphics*{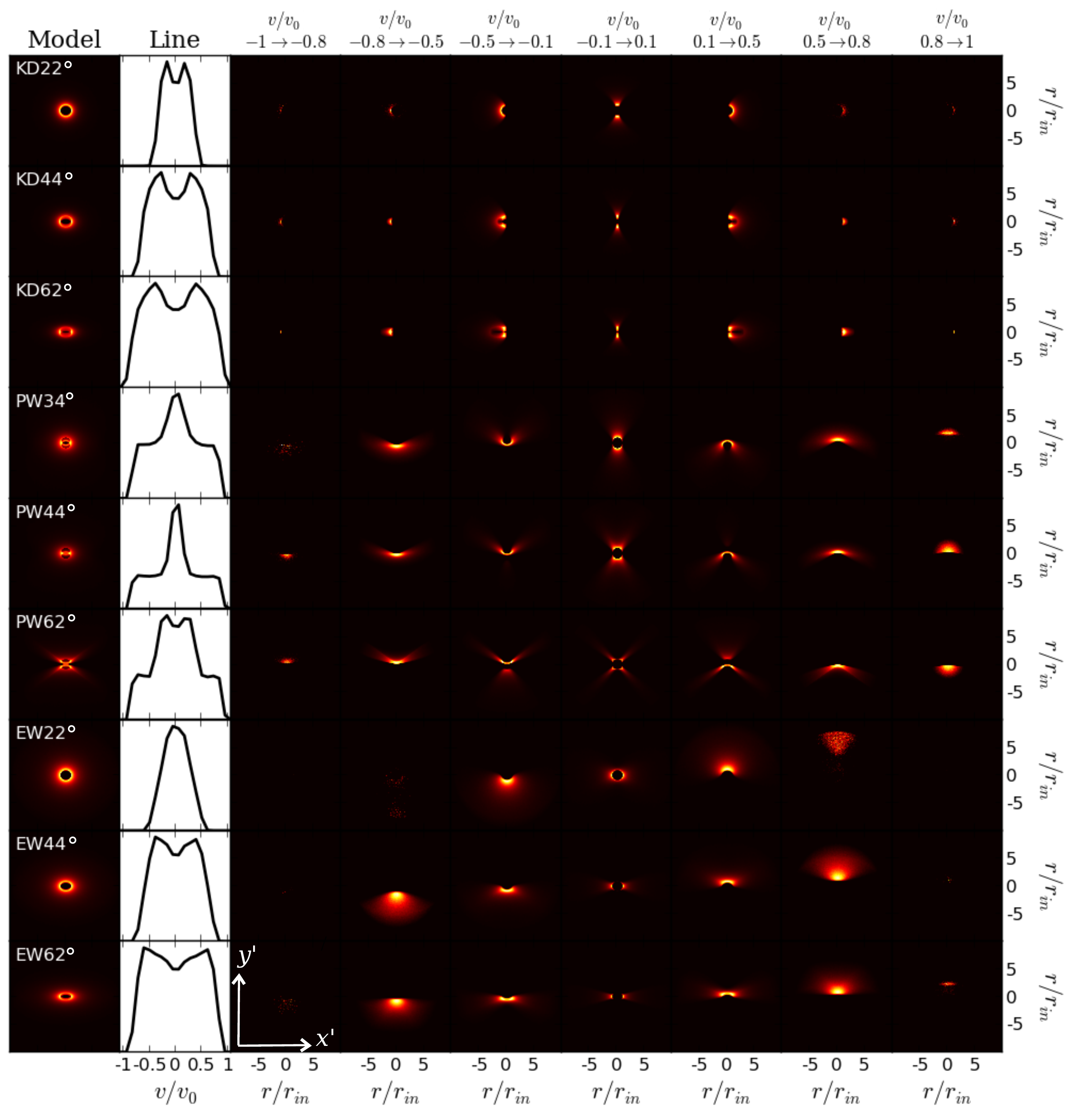}}
\caption{Emission from a thin Keplerian disk (KD), a biconical radially accelerated polar wind (PW), and a radially accelerated equatorial wind (EW) in projection onto the plane of the sky, for a set of representative inclinations of the line of sight to the observer. The inclination of the line of sight is quoted in degrees, next to the model initials. For each BLR model, Col.~1 is the wavelength-integrated surface brightness; Col.~2 is the emission line profile; and Cols.~3-9 are the surface brightness of the emission region that contributes to a given velocity range in the line profile.\ The parameter $v_0$ is the highest velocity of the BLR gas. In the Keplerian disk and equatorial wind, the maximum projected velocity that determines the width of the line at zero intensity decreases with decreasing inclination of the line of sight.}
\label{fig:BLR_models}
\end{figure*}

The surface brightness of the models as seen by detectors at different polar angles is illustrated in Fig.~\ref{fig:BLR_models}, integrated in seven narrow bands sampling the velocity structure of the line profile. Although simulations are performed over a symmetric velocity range, small differences between the surface brightness in corresponding slices of negative and positive velocities can be noticed owing to the asymmetric inclusion or exclusion of the limits of the velocity bins.

\subsubsection{Mean emission radius and direction}

Insight about the size and shape of the BLR, and in particular about the evolution of its extent and orientation as a function of wavelength, is achieved through the computation of the mean emission radius and direction. The mean emission radius, $\rho_m$, is defined as the radial distance weighted by the (monochromatic) surface brightness of the BLR seen in projection onto the plane of the sky,
\begin{equation}
\rho_m = \frac{\int_0^{r_{\text{out}}} \int_0^{2 \pi} \, I(\rho,\theta) \, \rho \,\text{d} \theta \, \text{d} \rho}{\int_0^{r_{\text{out}}} \int_0^{2\pi} \, I(\rho,\theta) \,\text{d} \theta \, \text{d} \rho} \hspace{2mm} ,
\label{eq:mean_em_radius}
\end{equation}
where $(\rho,\theta)$ are the polar coordinates in the plane of the sky and $I$ is the intensity of the emission from the BLR. The mean emission direction, $\theta_m$, is obtained similarly, i.e.,
\begin{eqnarray}
\cos \, \theta_m  & = &\frac{\int_0^{r_{\text{out}}} \int_0^{2\pi} \, I(\rho,\theta) \, \cos \, \theta  \, \text{d} \theta \,\text{d} \rho}{\int_0^{r_{\text{out}}} \int_0^{2\pi} \, I(r,\theta) \, \text{d} \theta \, \text{d} \rho} \hspace{5mm} , \nonumber \\
\sin \, \theta_m  & = &\frac{\int_0^{r_{\text{out}}} \int_0^{2\pi} \, I(\rho,\theta) \, \sin \, \theta  \,\text{d} \theta \, \text{d} \rho}{\int_0^{r_{\text{out}}} \int_0^{2\pi} \, I(\rho,\theta) \,\text{d} \theta \, \text{d} \rho} \hspace{2mm} .
\label{eq:mean_em_direction}
\end{eqnarray}

The mean emission radius and direction can be computed using either the BLR emission integrated over wavelength or the ``monochromatic'' emission. The latter allows the determination of the position of the emission region as a function of the Doppler velocity shift. The relative position of the low- and high-velocity regions of the BLR, and of its blueshifted and redshifted emission regions are of particular interest when exploring the possibility that microlensing can produce wings/core and red/blue differential magnification of the line profile.

\subsection{Characteristics of the adopted BLR models}
\label{sec:prop_blr}

\begin{figure*}
\centering
\resizebox{\hsize}{!}{\includegraphics*{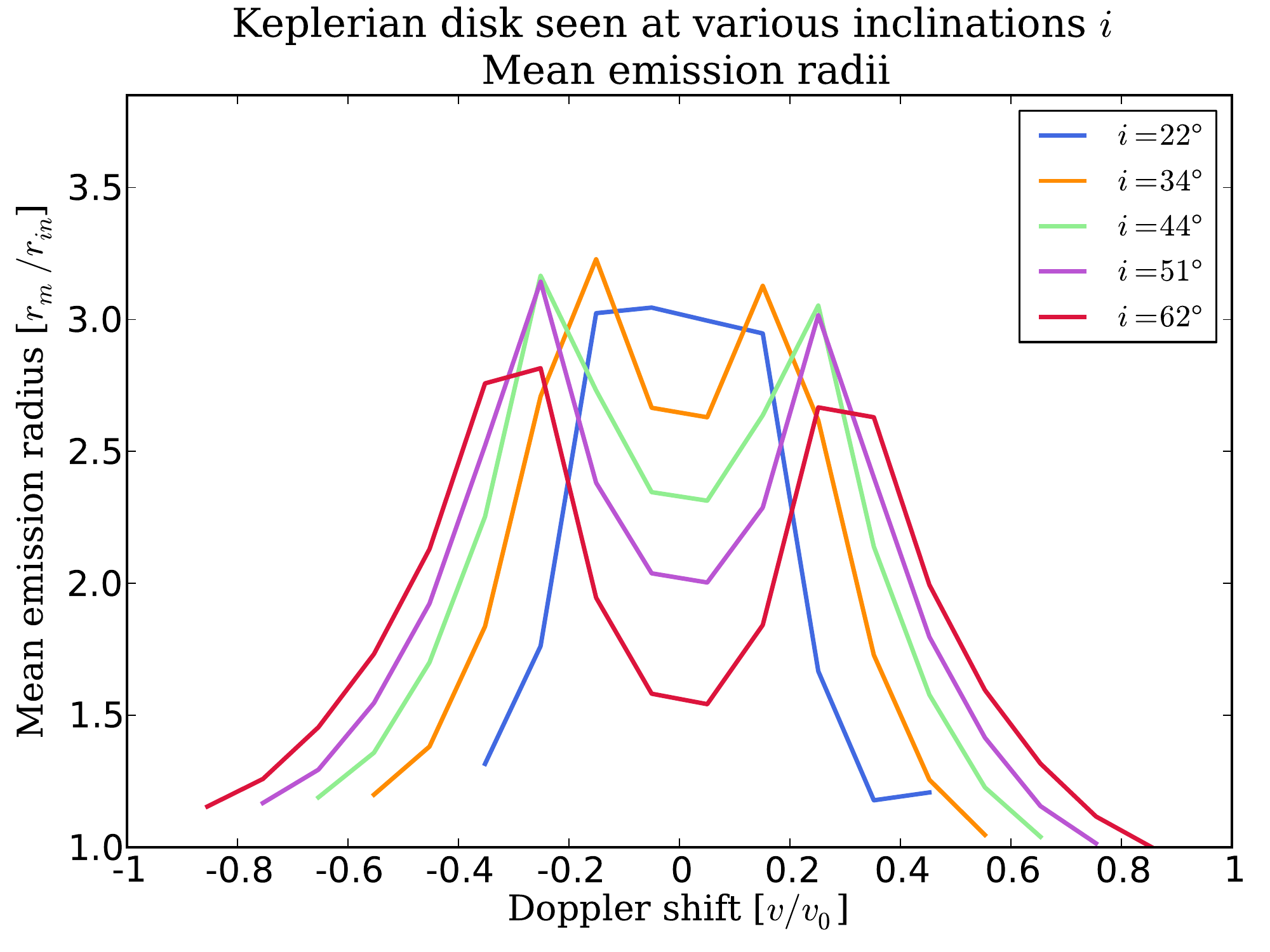}
  \includegraphics*{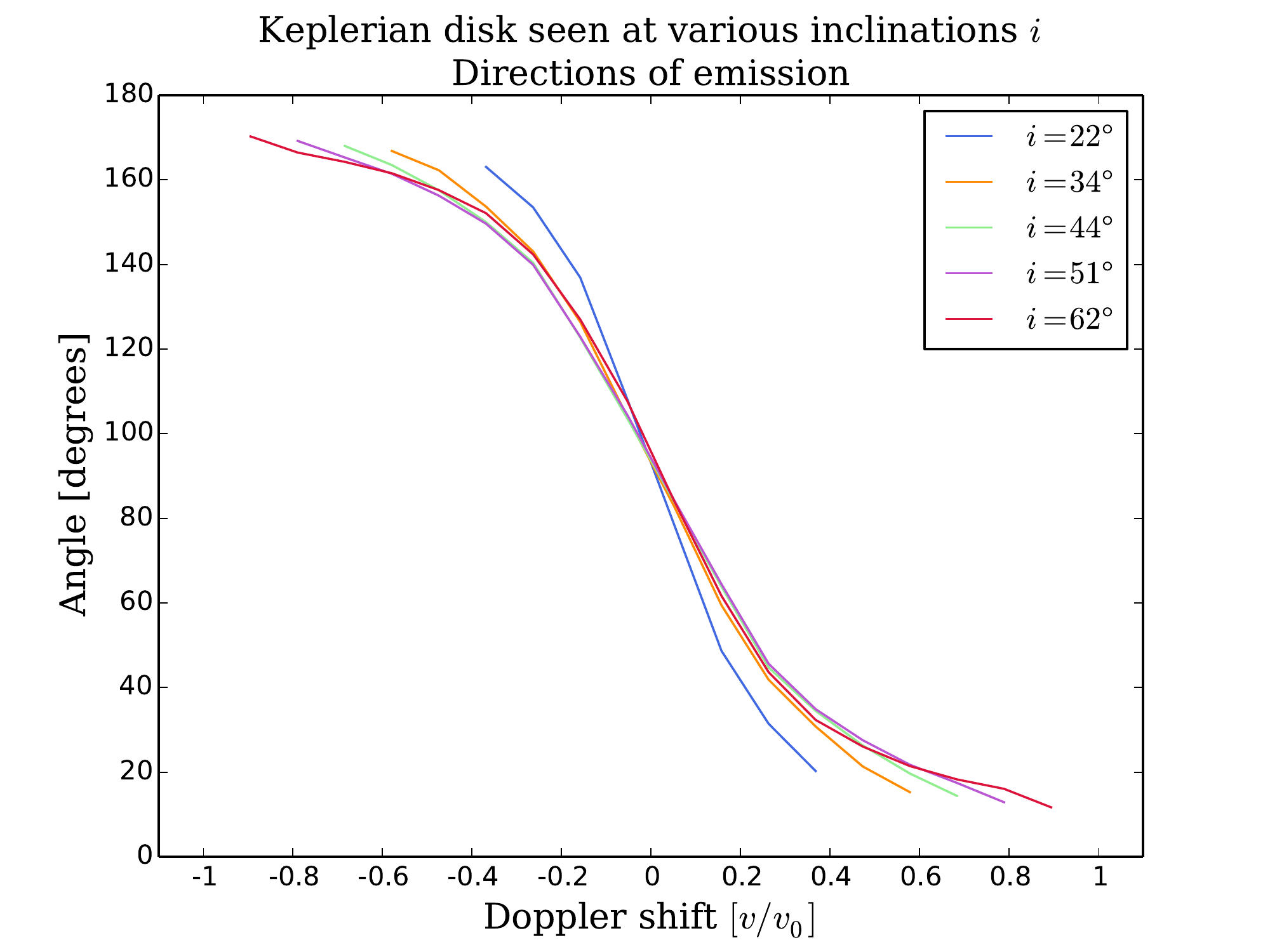}
  \includegraphics*{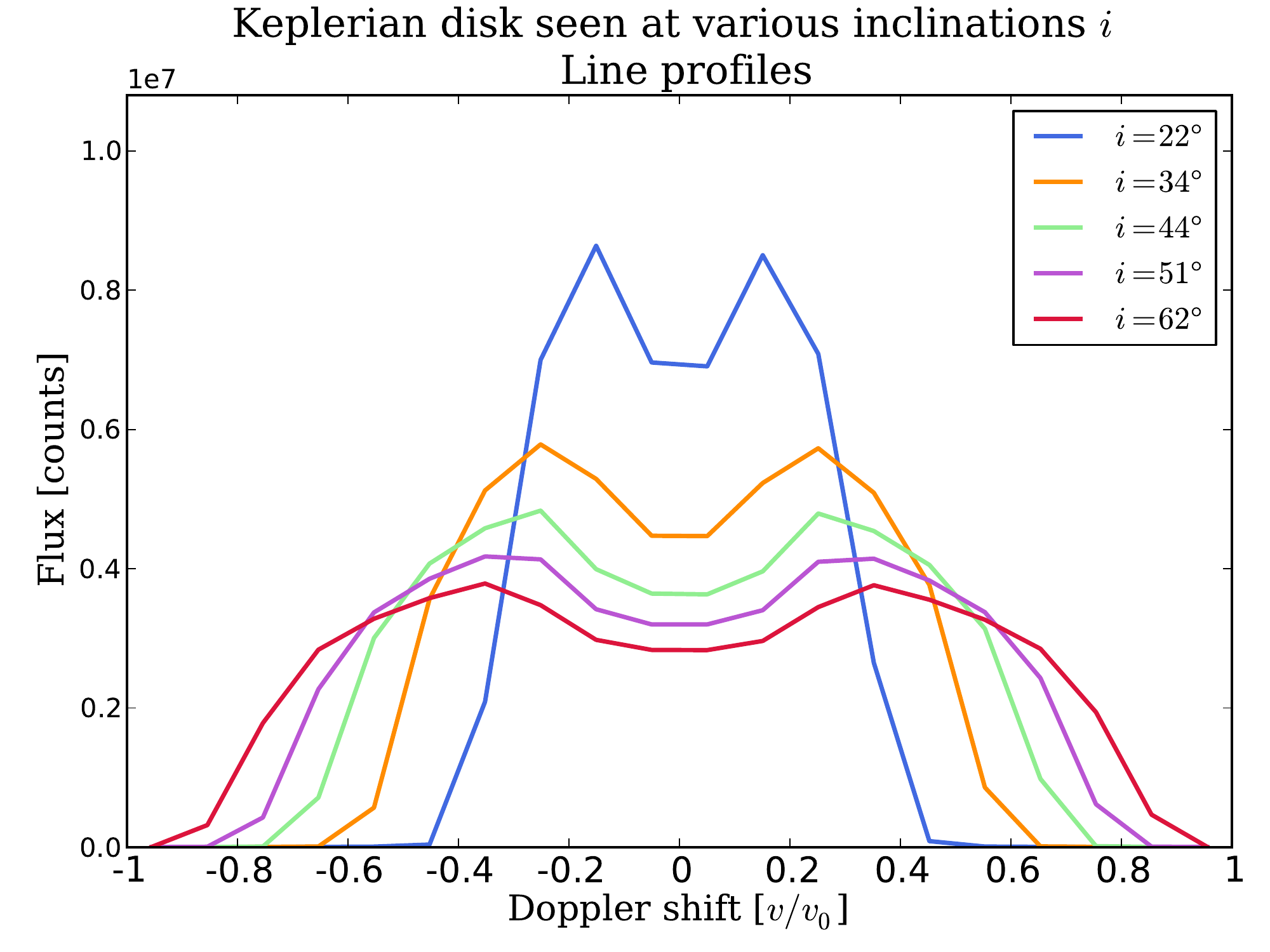}}
\label{fig:disk-lineprofile}
\caption{Keplerian disk model.
  \textit{Left:} The mean emission radius as a function of the Doppler velocity shift.
  \textit{Middle:} The mean emission direction relative to $x'$-axis as a function of the Doppler velocity shift.
  \textit{Right:} The broad emission line profiles.
  As explained in Sect.~\ref{sec:spaspesam}, the small red/blue asymmetries seen in the mean emission radii and the line profiles are not physical.}
\label{fig:disk-lineprofile+hlradius}
\end{figure*}
\begin{figure*}
\centering
\resizebox{\hsize}{!}{\includegraphics*{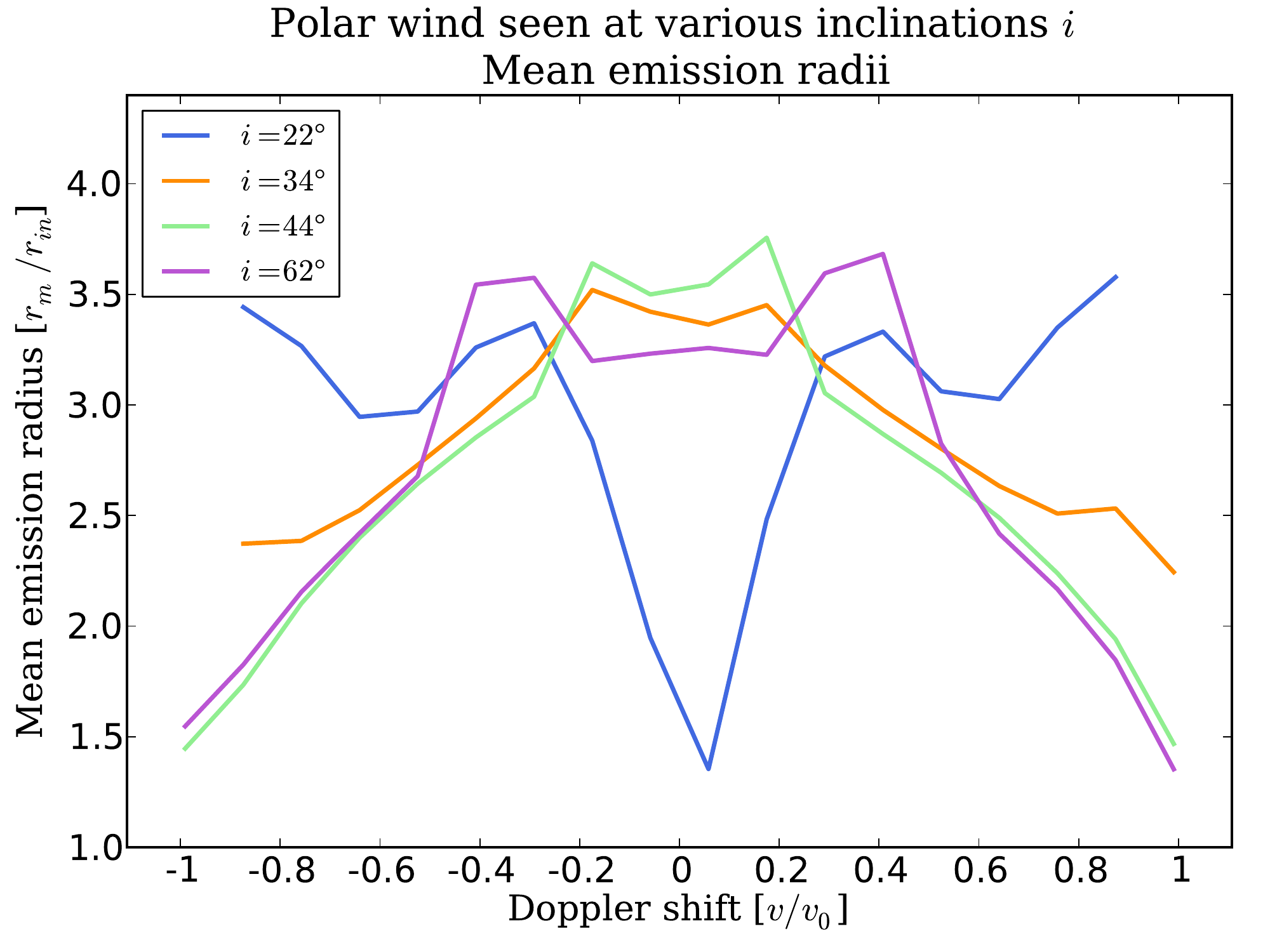}
\includegraphics*{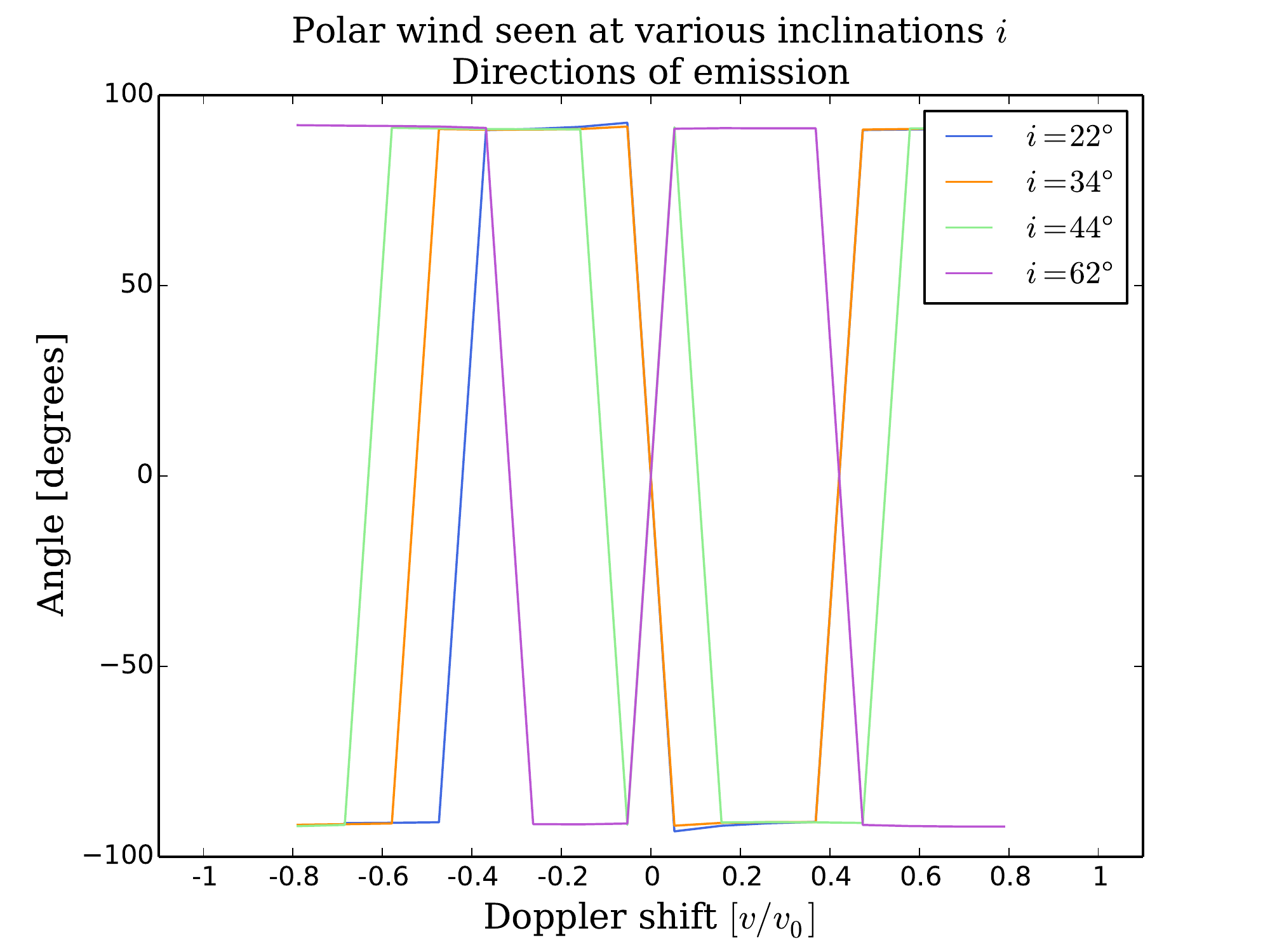}
\includegraphics*{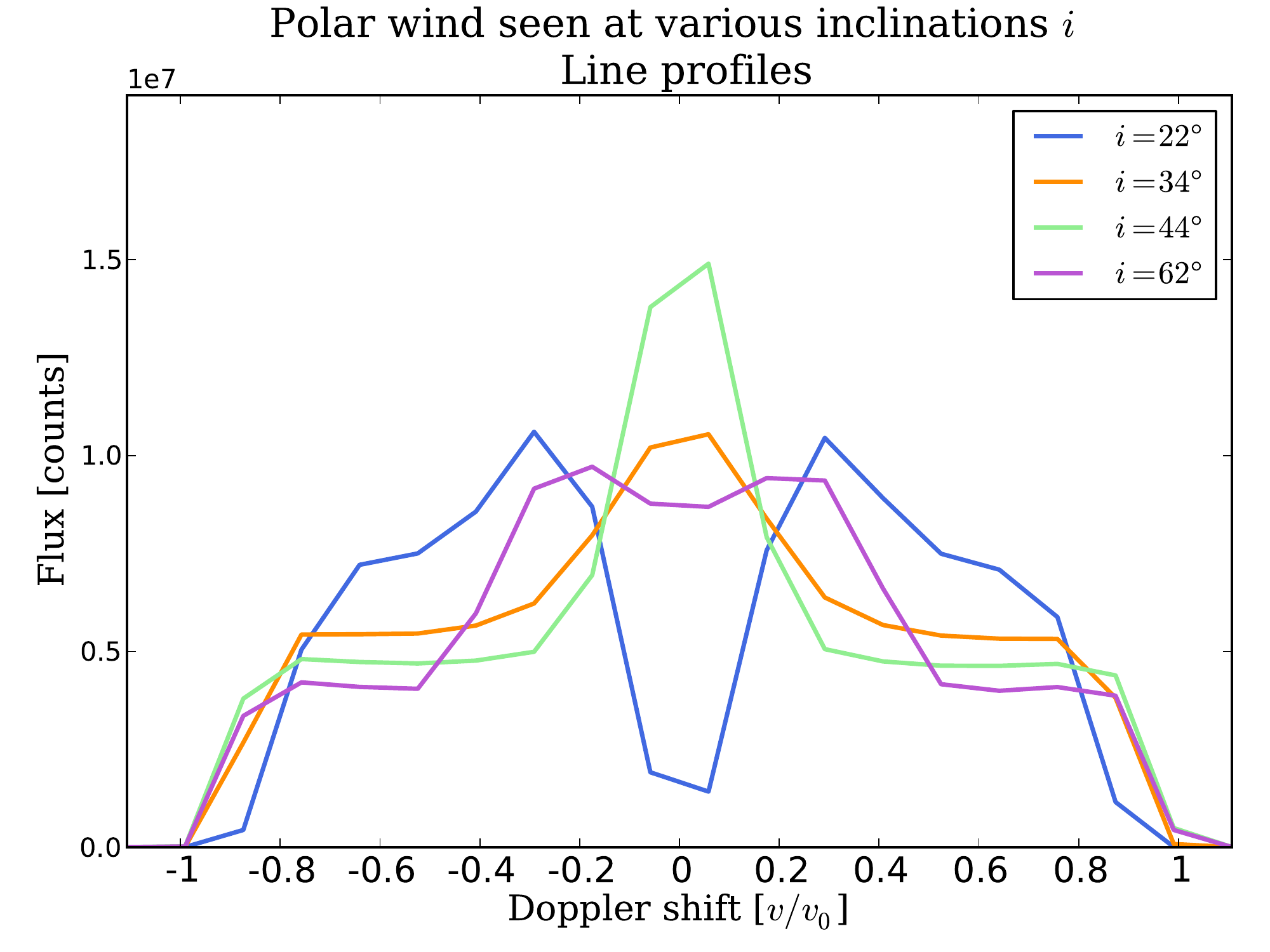}}
\caption{Same as Fig.~\ref{fig:disk-lineprofile+hlradius} for the polar wind model.}
\label{fig:polarwind-lineprofile+hlradius}
\end{figure*}
\begin{figure*}
\centering
\resizebox{\hsize}{!}{\includegraphics*{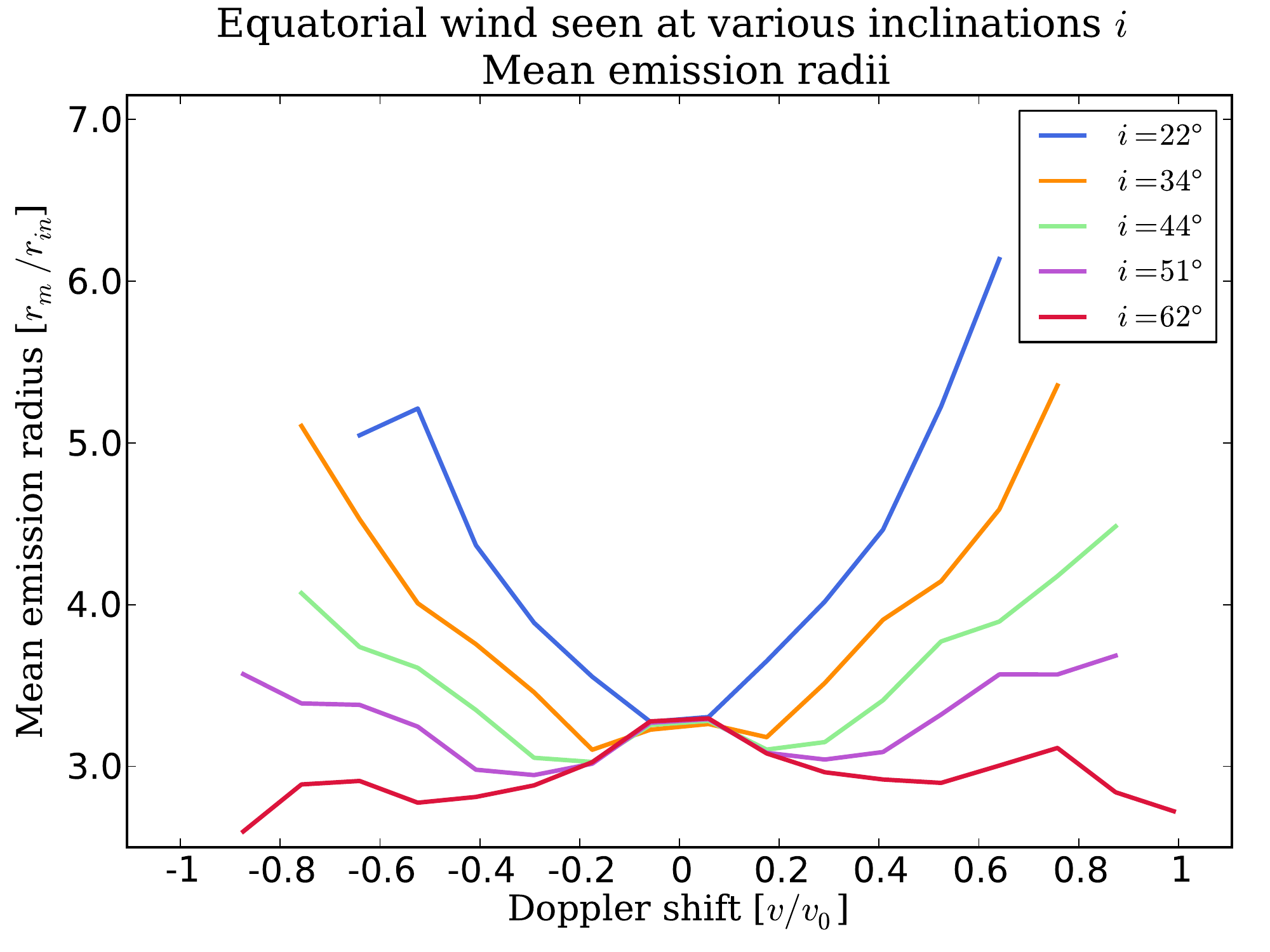}
\includegraphics*{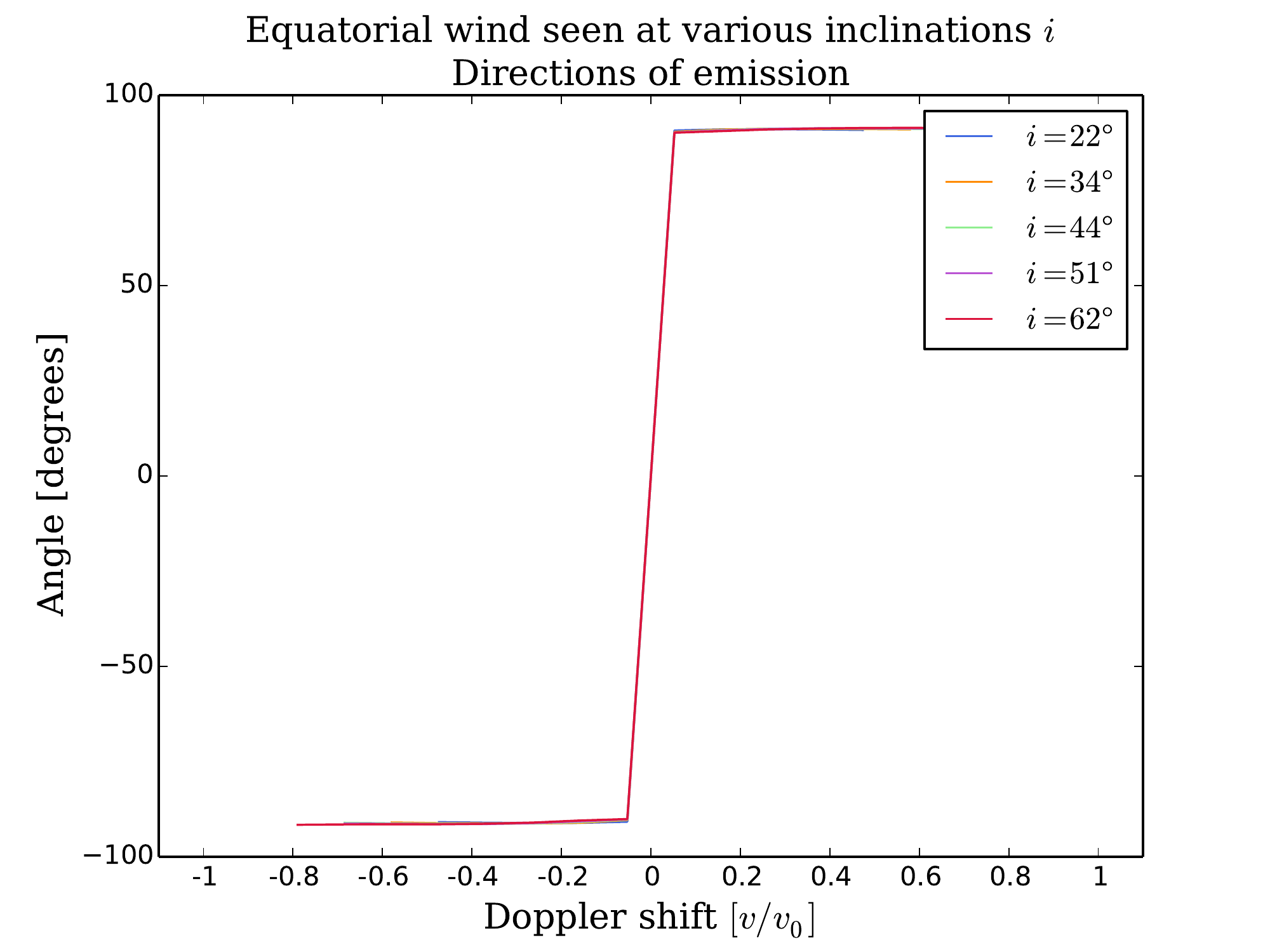}
\includegraphics*{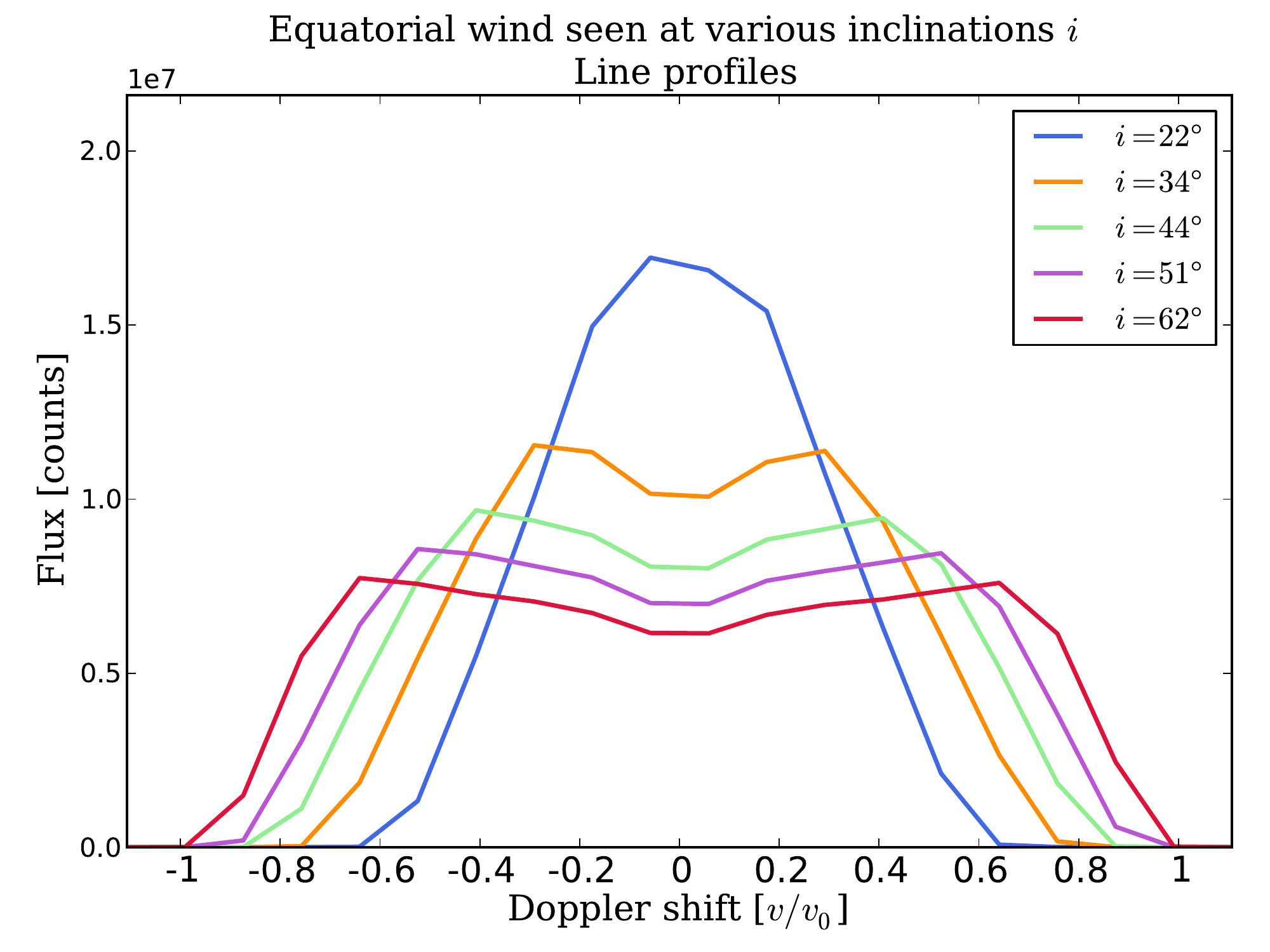}}
\caption{Same as Fig.~\ref{fig:disk-lineprofile+hlradius} for the equatorial wind model.}
\label{fig:equatwind-lineprofile+hlradius}
\end{figure*}

\subsubsection{Keplerian disk}
\label{sec:disk}

The thin Keplerian disk is defined as a cylinder (Fig.~\ref{fig:models_geometry}) that extends from $r_{\text{in}}$ to $r_{\text{out}} = 10 \, r_{\text{in}}$ and has a height of $1\,r_{\text{in}}$. It is animated by a pure azimuthal motion (Eq.~\ref{eq:vazim}). The appearance of the thin Keplerian disk for several inclinations of the line of sight relative to the polar axis is presented in Fig.~\ref{fig:BLR_models} (rows 1-3). The surface brightness of the models projected onto the plane of the sky is illustrated wavelength integrated and in seven narrow bands sampling the velocity structure of the line profile.

The inclination of the line of sight clearly controls the width of the emission line. Low-inclination (i.e., polar) lines of sight lead to smaller projected velocities. The inclination of the line of sight does not change the overall shape of the emission region seen at different Doppler velocity shifts. Indeed, regardless  the inclination, the low-velocity core of the line arises in a thin central region, while the blue and red line wings originate from opposite sides of the disk with the highest velocities coming from a narrow region, close to the center. 

The mean emission radius computed for the Keplerian disk model is significantly smaller than the mean emission radius of wind models (Table~\ref{tab:mean_emission_radius}). This indicates that a larger percentage of the light is emitted at small radii in the Keplerian disk. The variation of the mean emission radius and direction through the velocity structure of the line profile, computed using the BLR ``monochromatic'' surface brightness, is shown in Fig.~\ref{fig:disk-lineprofile+hlradius}. From negative to positive velocities, the mean emission direction smoothly ``switches'' from one side of the disk to the other. Moreover, in agreement with the more and more compact emission regions observed at high blue and red Doppler shifts in Fig.~\ref{fig:BLR_models}, the mean emission radius clearly diminishes at high velocities whatever the inclination. The mean emission radius also drops at very low Doppler shift in the line core. The magnitude of that drop depends on the viewing angle. The higher is the inclination of the line of sight, the smaller the mean emission radius of the low-velocity region. This is due to the projection of the circular emission region into an ellipse at high inclinations.

\subsubsection{Polar radially expanding wind}
\label{sec:polwind}

The polar wind is modeled as a biconical shell with an inner opening angle of $45\degr$ and an outer opening angle of $60 \degr$ (Fig.~\ref{fig:models_geometry}). Its velocity field is a pure radial outflow that accelerates outward following Eq.~\ref{eq:vradial}.

The line profile that arises from the biconical outflow displays an important lack of flux in its core when the polar wind is seen at low inclination (Fig.~\ref{fig:polarwind-lineprofile+hlradius}). The missing low-velocity line flux is due to the nonzero velocity of the gas at the base of wind (Eq.~\ref{eq:vradial}) in combination with an inclination that is not sufficient to produce null projected velocities. Besides, the low-velocity region is confined to the innermost radii of the accelerating outflow when it is seen at low inclination (i.e., pole-on), so that the drop in the line core corresponds to a net decrease of the mean emission radius. Such a lack of flux in the line core prevents an accurate detection of possible wings/core effects. Investigations of the BLR polar wind model are therefore restricted to lines of sight with $i \geq 30 \degr$. 

As for the Keplerian disk, the mean emission radius of the polar wind changes with the Doppler velocity shift but the variation is smaller (Fig.~\ref{fig:polarwind-lineprofile+hlradius}), which implies a smaller difference of compactness between the slow and fast parts of the flow (seen in projection). The polar wind emission (Fig.~\ref{fig:BLR_models}, rows 4-6) appears to originate from a more extended region than the Keplerian disk, both in integrated and monochromatic line flux, in agreement with the larger mean emission radius reported in Table~\ref{tab:mean_emission_radius}.

In contrast with the Keplerian disk and the equatorial wind (Sect.~\ref{sec:eqwind}), the approaching and receding parts of the velocity field (with respect to the observer) appear partly co-spatial in projection, especially for low and intermediate velocities. This explains the multiple switches of the mean emission direction between $-90\degr$ to $+90\degr$ within the range of positive (respectively, negative) Doppler shifts (Fig.~\ref{fig:polarwind-lineprofile+hlradius}). Indeed, since the radially outflowing velocity field shows no abrupt velocity change but evolves smoothly, those sudden variations of the mean emission direction result from a change in the contrast between the brightness of the lower (i.e., $\theta_m=-90\degr$) and upper (i.e., $\theta_m=+90\degr$) parts of the wind.

\begin{table}[t]
\centering
\caption{Mean emission radius $\rho_m/r_{\text{in}}$ of the three BLR models for some representative values of the inclination of the line of sight, considering a rapidly decreasing emissivity $\epsilon(r) = \epsilon_0 \,  (r_{\text{in}}/r)^{3}$.}
\label{tab:mean_emission_radius}
\begin{tabular}{l|ccccc}
{} & $i=$ & $22\degr$ & $34\degr$ & $44\degr$ & $62\degr$ \\
\hline
Keplerian disk  & & $2.47$ & $2.35$ & $2.23$ & $1.96$ \\
Polar wind      & & $3.06$ & $3.01$ & $2.98$ & $3.04$ \\
Equatorial wind & & $3.73$ & $3.54$ & $3.37$ & $2.97$ \\
\end{tabular}
\end{table}

\subsubsection{Equatorial radially expanding wind}
\label{sec:eqwind}

The third model considered is an equatorial radially accelerating  wind that has a $30 \degr$ opening angle (Fig.~\ref{fig:models_geometry} and Eq.~\ref{eq:vradial}).

As in the Keplerian disk model, the equatorial wind exhibits a clean spatial separation between the negative and positive parts of its velocity field (Fig.~\ref{fig:BLR_models}, rows 7-9). However, because of the radial acceleration of the wind, the gas flows faster at larger distance. Consequently, unlike virialized velocity fields, the mean emission radius increases at higher Doppler shifts in equatorial winds. Highly approaching and receding velocities are found in the outer parts of the wind, located in opposite directions, so that the bluest and reddest parts of the line come from regions of the wind that are clearly spatially separated. The switch in the mean emission direction between negative and positive Doppler shifts is much sharper in the equatorial wind than in the Keplerian disk. This is because the emission occurs at radii larger than $\sim 3\, r_{\text{in}}$ in the equatorial wind, at every Doppler shift and whatever the inclination (Fig.~\ref{fig:equatwind-lineprofile+hlradius}), thus at much larger radii than in the Keplerian disk (Table~\ref{tab:mean_emission_radius}). Since the emission direction is better constrained when emission arises at large distance than when most emission is radiated close to the center, the variation of $\theta_m$ appears sharper for the equatorial wind model.

\section{Modeling the effect of microlensing on the BLR}
\label{sec:modelingMicrolensing}

\subsection{Generic caustic patterns}
\label{sec:method_to_simulate_microlensing1}

Microlensing is modeled by a Chang-Refsdal lens \citep{1979ChangRefsdal,1984ChangRefsdal}, which describes the magnification caused by a single star perturbed by the tidal field of the lensing galaxy to which it belongs. The simple patterns produced by the Chang-Refsdal lens provide a relevant representation of the magnification by ``fold'' and ``cusp'' caustics. These patterns appear to be a good compromise between the complexity of a realistic network of caustics that varies from object to object and the practicability offered by idealized patterns. They are therefore well suited to investigate whether the effect of microlensing on the line profiles can lead to the discrimination between the different models of BLR.

Chang-Refsdal magnification patterns are characterized by two parameters: the surface mass density $\kappa$ and the shear $\gamma$ of matter at the location of the microlensing star. The Einstein radius $r_E$
\begin{equation}
  r_E = \sqrt{4 \; \frac{GM}{c^2} \frac{D_S D_{LS}}{D_L}}
\end{equation}
characterizes the cross section of gravitational lensing, i.e., the area of the source that can be significantly magnified by a lens of mass $M$. Parameters $D_S$, $D_L$ and $D_{LS}$ are the source, lens, and lens-source angular diameter distances, respectively. The Einstein radius is the natural scale length in which the dimensions of magnification maps are expressed. A solar-mass microlens in a typical gravitational lens system with the lens and source located at redshifts $z_L = 0.5$ and $z_S = 2.0$, has an Einstein radius $r_E \simeq 20 \sqrt{M/M_{\odot}} \,$light days $\simeq 0.017 \sqrt{M/M_{\odot}}$ pc, adopting a flat $\Lambda$CDM cosmology with $H_0 = 68$ km s$^{-1}$ Mpc$^{-1}$ and $\Omega_m$ = 0.31.

\subsection{Effect of microlensing on the BLR}
\label{sec:method_to_simulate_microlensing2}

\begin{figure}
\centering
\resizebox{\hsize}{!}{\includegraphics*[trim={0 0 0 17mm},clip]{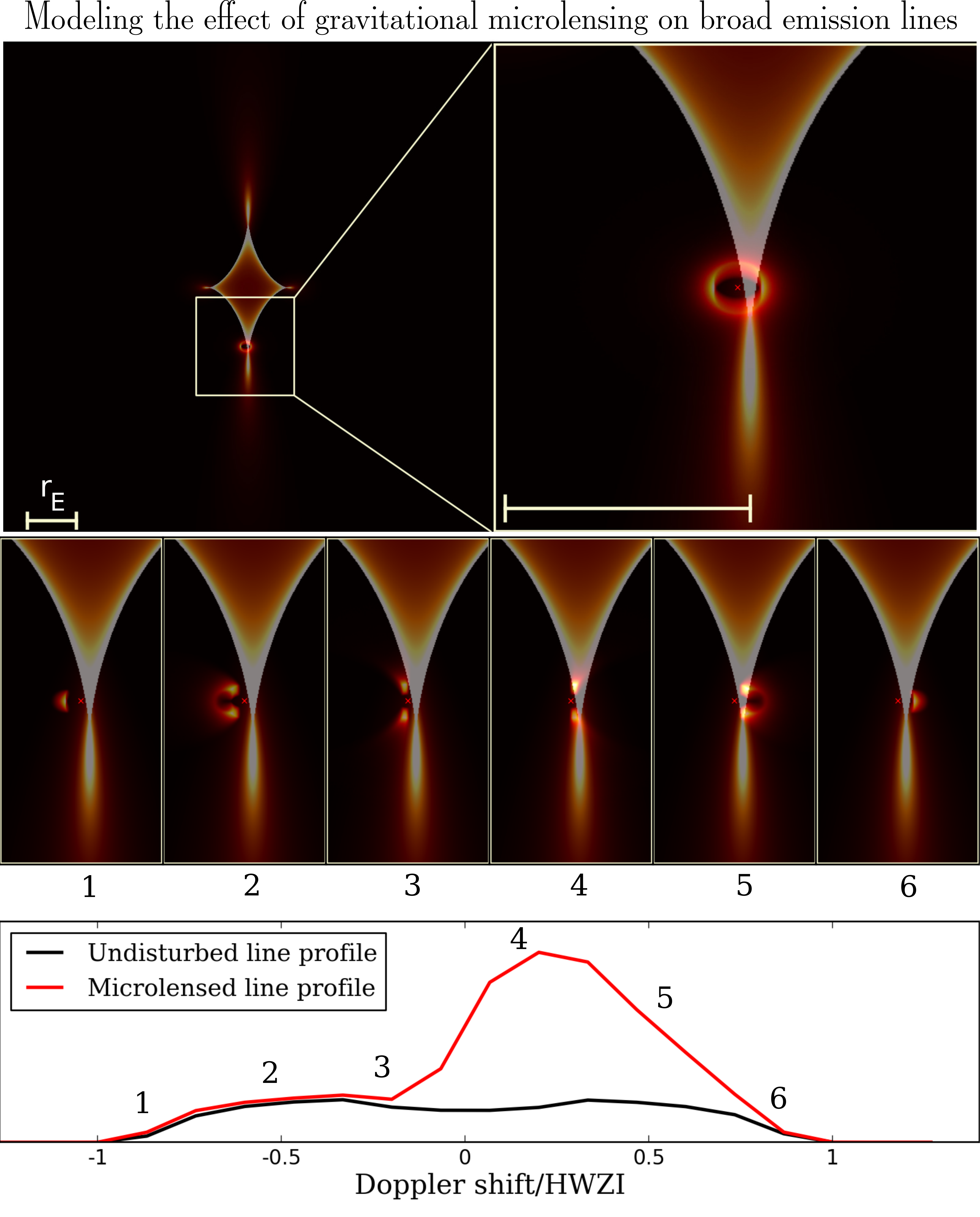}}
\caption{Effect of gravitational microlensing on broad line emission. The illustrated caustic structure is computed for a $(\kappa,\gamma)=(0,0.5)$ Chang-Refsdal lens. The BLR is a Keplerian disk seen nearly face-on, whose size is fixed to $r_{\text{in}} = 0.1\, r_E$. The Einstein radius, $r_E$, is indicated in the upper panels. The upper panels show the emission of the Keplerian disk, integrated over the line profile, and superimposed on the magnifying structure. The upper right panel is a zoom of a $2 \, r_E \times 2 \, r_E$ region. The intermediate series of panels present snapshots of the BLR emission in six bins of Doppler velocities sampling the line profile, superimposed on the magnification map. The microlensed flux is obtained by the multiplication of the snapshots of the BLR surface brightness at different Doppler velocities with the region of the caustic structure sampled by the BLR model. The resulting distorted line profile is plotted as a solid red line in the lower panel.}
\label{fig:modellingBLR-principle}
\end{figure}

To simulate distortions of line profiles, the microlensed line flux is computed in each spectral bin of the line profile as the product of the part of the BLR emission that contributes to the considered range of Doppler velocities with the magnification pattern computed in the source plane, over the spatial area that encapsulates the BLR. The BLR indeed covers only a fraction of the magnification map (Fig.~\ref{fig:modellingBLR-principle}). The size of the BLR model is settled with respect to the extent of the caustic structure by expressing the inner radius of the BLR, $r_{\text{in}}$, which is the reference length scale of the BLR model in terms of the Einstein radius, $r_E$, which is the reference length scale of the caustic structure. The BLR model is spatially resampled to match the spatial sampling of the magnification map. Considering a BLR element that has a surface brightness $I(x',y',\lambda)$ in the spectral bin centered on $\lambda$ and that is located at a position $(X,Y)$ in the magnification map $\mu(X,Y)$ in the source plane, the microlensed line flux $F(X,Y,\lambda)$ that results from the magnification of that ``monochromatic snapshot'' of the BLR emission is given by
\begin{equation}
F(X,Y,\lambda) = \int \int I(x',y',\lambda) \, \mu(X-x',Y-y') \, \text{d}x' \text{d}y'
\label{eq:convolution}
,\end{equation}
where $(X,Y)$ is the coordinate system associated with the magnification map in the source plane and $(x',y')$ the coordinate system in the sky plane onto which the BLR is projected. Practically, the microlensed line flux, and thus the distorted line profile, is computed at each point of the magnification pattern as the convolution of the BLR surface brightness with the magnification map (Eq.~\ref{eq:convolution}). This allows us to explore the effect of microlensing on the line profile for all positions of the BLR model onto the caustic structure.

\begin{figure}
\centering
\resizebox{\hsize}{!}{\includegraphics*{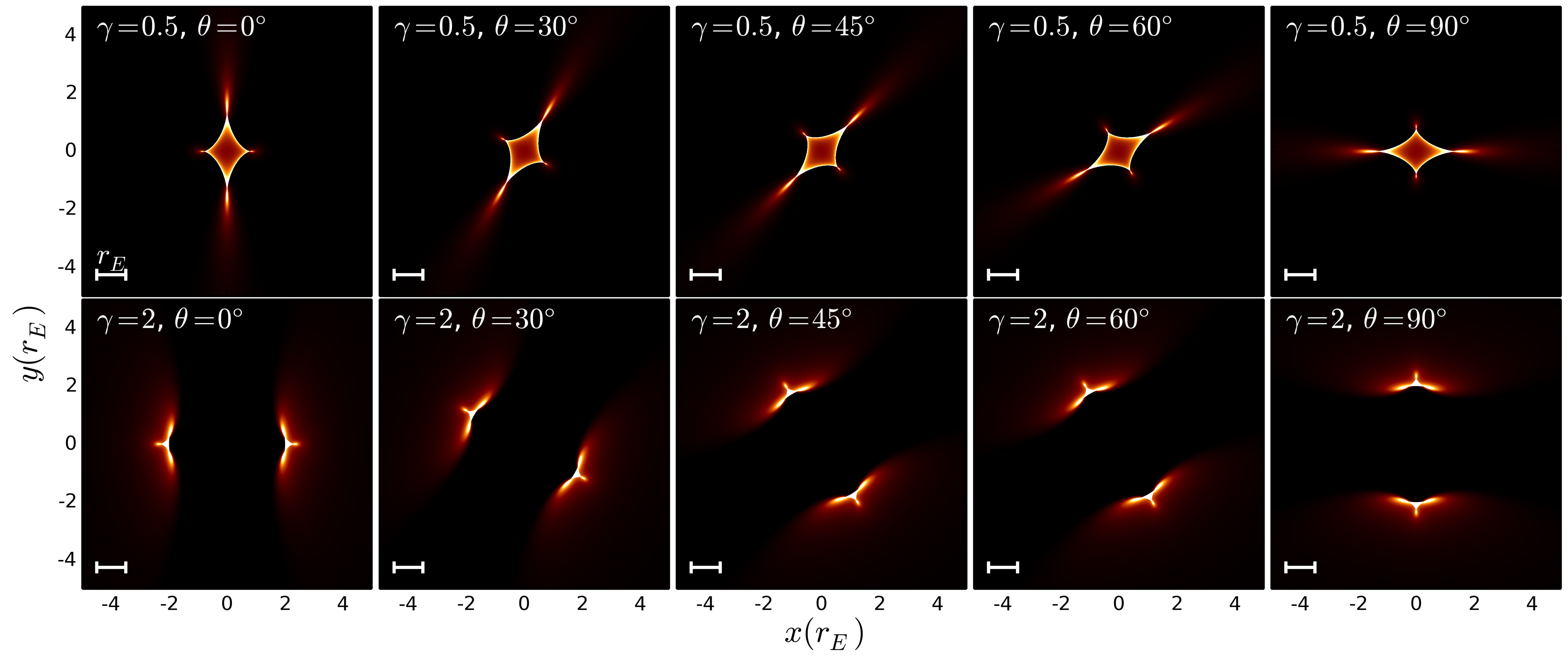}}
\caption{Magnification maps obtained for a $(\kappa=0,\gamma=0.5)$ Chang-Refsdal lens (upper panels) and a $(\kappa=0,\gamma=2)$ Chang-Refsdal lens (lower panels). From left to right, the caustics are rotated by $0\degr$, $30\degr$, $45\degr$, $60\degr$, and $90\degr$ to prevent biases due to alignments between the BLR model symmetry axes and the caustics. The magnification maps are simulated over a $30  \, r_E\times 30 \, r_E$ region of the source plane but the above panels illustrate a $10  \, r_E \times 10 \, r_E$ zoom of the caustic structure. }
\label{fig:caustic_maps}
\end{figure}

We build $30  \, r_E \times 30 \, r_E$ magnification maps sampled by $6000 \times 6000$ pixels. We  assume $\kappa=0$. Depending on the shear, $\gamma$, the Chang-Refsdal lens produces either a ``diamond-shaped'' astroid caustic or a pair of ``triangle-shaped'' deltoid caustics. Magnification maps were computed for $\gamma = 0.5$ and $\gamma = 2.0$. Both caustic configurations are investigated. The caustic maps are rotated by $30\degr$, $45\degr$, $60\degr$, and $90\degr$ to ensure that the study is not biased by  alignments with the symmetry axes of the BLR models (Fig.~\ref{fig:caustic_maps}). 

The size of the BLR ranges from a few up to several hundreds light days \citep{2000Kaspi,2005Kaspi,2007Kaspi,2013Guerras}. Accounting for the range of Einstein radii in gravitational lenses \citep{2011MosqueraKochanek}, the smallest BLR size is on the order of $0.1 \, r_E$. We thus investigate nine inner radius values for each BLR model: $r_{\text{in}}= \, 0.1, \, 0.125, \, 0.15, \, 0.175, \, 0.2, \, 0.25, \, 0.35, \, 0.5,$ and $\, 0.75 \, r_E$ with $r_{\text{out}}=10 \, r_{\text{in}}$. No significant microlensing effect is expected for larger BLRs \citep{1991RefsdalStabell}. Given that the mean emission radius of the BLR models is $\rho_m \sim 3 \, r_{\text{in}}$ (Table~\ref{tab:mean_emission_radius}), this corresponds to a mean emission radius ranging from about $0.3$ to $2.3 \, r_E$.

\section{Modeling the effect of microlensing on the continuum source}
\label{sec:continuum_source}

Differential microlensing in the BELs is commonly observed in systems also showing microlensing of the continuum \citep{2012Sluse}. Since the accretion disk at the origin of the continuum emission and the BLR are microlensed by the same caustic structure, the magnification experienced by the continuum thus provides an important additional constraint on the position of the system with respect to the caustic pattern. As for the BLR models, the microlensing magnification experienced by the continuum source is computed at each position on the magnification map by convolving this map with the surface brightness of the accretion disk so that each pixel of the resulting convolved map holds the microlensed continuum emission. That microlensed flux is then divided by the initial flux in the continuum to obtain the magnification of the continuum $\mu^{cont}$.

The accretion disk is modeled as a disk of constant surface brightness (uniform disk). \citet{2005Mortonson} showed that the effect of microlensing on circular disk models is rather insensitive to their surface brightness profile. For a wide range of disk radii, exceding those considered here, differences in magnification do not exceed a few percent. The half-light radius $R_{1/2}$ of the continuum source is indeed found to be the primary parameter that controls the amplitude of the magnification by the caustic structure, i.e., the smaller the half-light radius the higher the magnification.

Based on microlensing studies, the measured half-light radii of accretion disks range from $0.75$ to $59.4 \,$light days (after rescaling to a mean microlens mass $M$ = $0.3 \, M_{\odot}$), most of which are between $2$ to $15 \,$light days \citep{2010Morgan,2011Blackburne,2014Jimenez}. Considering that the Einstein radius in lensed systems varies between $5$ and $40 \,$light days for $M$ = $0.3 \, M_{\odot}$ \citep{2011MosqueraKochanek}, continuum sources have half-light radii between $0.05 \, r_E$ and $3 \, r_E$. Specifically, we investigate uniform disks with outer radii $r_s = \sqrt{2} \, R_{1/2}$ fixed at $9$ different values: $r_s = 0.1, \, 0.15, \, 0.2, \, 0.25, \, 0.3, \, 0.35, \, 0.4, \, 0.5, \,$ and $ 0.6\, r_E$. 

The accretion disk and the BLR model are assumed to share the same symmetry axis and their centers are assumed to coincide. Microlensing effects are therefore simulated for an accretion disk seen under the same inclination as the BLR model and magnified by the same caustic structure. Finally, the resulting magnification of the continuum and the deformations of the line profiles are only computed for realistic models in which the BLR is larger than the continuum source, i.e., $r_{\text{in}} \geq r_s$.

A larger emission region causes a higher ``smoothing'' of the caustic pattern after convolution. As a result, the maximum magnification experienced by the continuum decreases when the size of its source increases. In particular, microlensing magnification that is larger than two cannot be reproduced for disk sizes $r_s > r_E$, whatever the magnification pattern. Furthermore, larger magnifications are measured for accretion disks seen at high $i$, i.e., ``edge-on'', owing to the smaller projected area of inclined disks.

\section{Characterizing line profile microlensing: Definition of observables}
\label{sec:measure_distortions}

Given the huge amount of simulated line profiles, we focus our analysis on quantities that characterize the line profile magnification and distortions and that can be directly and quantitatively compared to observations.

The total microlensing-induced magnification of the emission line, $\mu^{BLR}$, is obtained as the ratio between the microlensed line flux and the non-microlensed line flux, after subtraction of the local microlensed and non-microlensed continua, respectively. We thus write
\begin{equation}
  \mu^{BLR} = \frac{\int_{v_{-}}^{v_{+}} \, F^l_{\mu} \, (v) \, dv}{ \int_{v_{-}}^{v_{+}} F^l \, (v) \, dv} \hspace{0.5cm} ,
\label{eq:cmp-mublr}
\end{equation}
where $F^l_{\mu}(v)$ and $F^l(v)$ denote the continuum-subtracted microlensed and non-microlensed flux densities, respectively, i.e., $F^l_{\mu}(v) = F_{\mu}(v) - F^c_{\mu}(v)$ and $F^l(v) = F(v) - F^c(v)$,  $F^c_{\mu}(v)$ and $F^c(v)$ representing the microlensed and non-microlensed flux densities of the underlying continuum, respectively. The limits
$v_{-}$ and $v_{+}$  are the lowest and highest velocities of the line profile, respectively. For the simulated line profiles, $v_{-} = - v_{+}$.

Line profile distortions result from a differential microlensing magnification of the BLR velocity structure. Distortion indicators are built based on the magnification, $\mu (v)$, undergone by the line at every Doppler velocity, $v$. \ The quantity $\mu (v)$ is computed as the ratio between the microlensed and the non-microlensed emission line profiles, after subtraction of the local microlensed and non-microlensed continua, respectively, i.e., $\mu (v) = F^l_{\mu}(v) / F^l(v)$.

Asymmetric line profile distortions arise from differential magnification of the blue and red parts of the emission line. The ratio between the magnification measured at corresponding negative and positive Doppler velocities is therefore exploited to build a ``red/blue indicator'', $RBI$, i.e.,
\begin{equation}
RBI = \frac{\int_0^{v_{+}} \log \left( \mu(v) \right) dv}{\int_0^{v_{+}} dv} - \frac{\int_{v_{-}}^0 \log \left( \mu(v) \right) dv}{\int_{v_{-}}^0 dv}\hspace{5mm} .
\label{eq:bluered_param}
\end{equation}
$RBI > 0$ indicates that the red part of the line profile undergoes, on average, a larger magnification than the blue one. 

The ability of microlensing to magnify differently the wings and core of the line constitutes another interesting characteristic. A ``wings/core indicator'', $WCI$, is designed using the ratio between the magnification experienced by the line profile at a Doppler velocity $v$, and the magnification at zero velocity,
\begin{equation}
WCI = \frac{\int_{v_{-}}^{v_{+}} \mu(v)/ \mu(v=0) \; dv}{\int_{v_{-}}^{v_{+}} dv} \hspace{5mm}.
\label{eq:wingcore_param}
\end{equation}
$WCI > 1$ indicates that the high-velocity part of the line profile (the wings) is, on average, more microlensed than its low-velocity part (the core).

\begin{figure}
\centering
\resizebox{\hsize}{!}{\includegraphics*{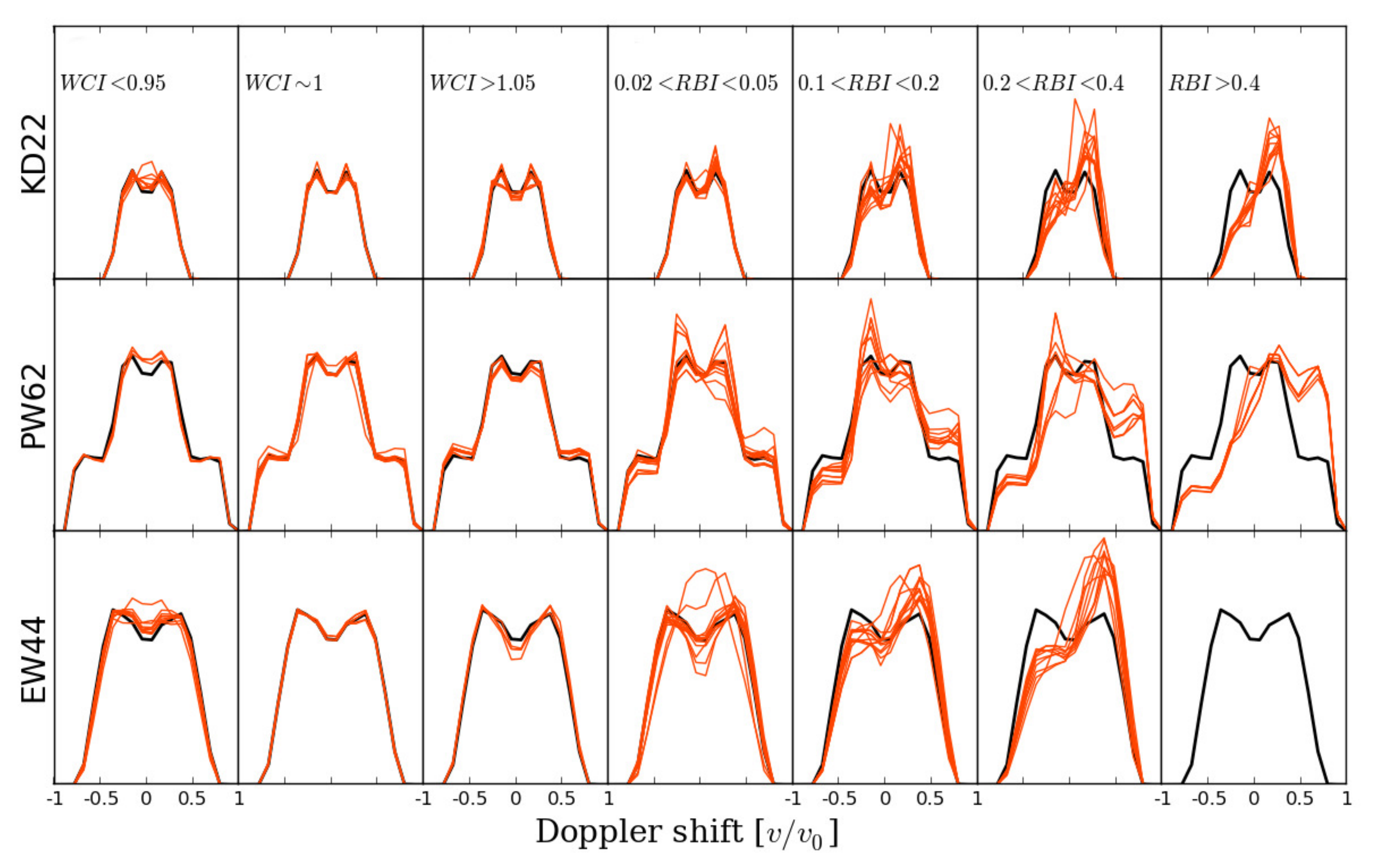}}
\caption{Examples of microlensed line profiles for different BLR models magnified by a   $(\kappa=0,\gamma=0.5)$  Chang-Refsdal caustic rotated by $0\degr$, $30\degr$, $45\degr$, $60\degr$, and $90\degr$. The microlensed line profiles (in orange) are normalized to the surface of the non-microlensed line profile (in black). The corresponding values of $WCI$ and $RBI$ are indicated. Some BLR models (EW44$\degr$) cannot produce large red/blue asymmetries ($RBI > 0.4$).}
\label{fig:profiles}
\end{figure}

Examples of microlensed line profiles and the corresponding distortions indicators are illustrated in Fig.~\ref{fig:profiles}.

The magnification $\mu (v)$ measured at different Doppler velocities, and consequently the distortion indicators, do not depend on the profile of the line, since $\mu (v)$ is defined as a ratio of line profiles.  This is of particular importance since our BLR models are not tuned to reproduce the line profiles observed in quasar spectra exactly. These indices, independent of the exact line profiles, focus on their distortions, and are thus particularly useful to compare microlensing simulations to observations.

We stress that BELs show intrinsic profile variations. Such variations usually follow a change of the continuum flux and propagate through the line profile with a time lag. This is at the basis of velocity-resolved reverberation mapping (Sect.~\ref{sec:intro}). Typical timescales are week to months, increasing with the AGN luminosity \citep[e.g.,][]{2010Bentz}. Line profile variations not related to reverberation effects are also reported on timescales of years \citep[e.g.,][]{1999Peterson,2010Lewis,2015Ilic}. In a lensed quasar, variations induced by microlensing magnification can thus be confused with intrinsic variations since the spectra of the different quasar images reach the observer with different time delays.  For luminous lensed quasars with small time delays, confusion with intrinsic variations is negligible, especially when sufficiently large profile distortions are considered \citep[][Appendix A]{2012Sluse}. For lensed quasar systems with large time delays (i.e., larger than 50 days), an adequate observational strategy is necessary; spectra of the different images must be obtained at two epochs separated by their respective time delays. This allows one to disentangle microlensing-induced from intrinsic variations, if the magnification $\mu (v)$ is computed using the spectra of two images obtained at epochs separated by the time delay.

\section{Results: General trends}
\label{sec:modelingBLR-results}

As discussed in the previous sections, microlensed line profiles are computed for each position of the BLR on the two magnification maps for the various adopted BLR models, inclinations, and radii. The continuum  magnification is also computed for several values of the accretion disk radius and considering the same inclinations as those of the BLR. Line profile distortions were compared to those of \cite{2002Abajas, 2007Abajas} and \cite{2004LewisIbata} and found in good qualitative agreement for models based on comparable assumptions.

The effect of microlensing is assessed with a set of four observables: $\mu^{BLR}$, the magnification experienced by the broad emission line;  $\mu^{cont}$, the magnification of the underlying continuum emission; as well as the red/blue, $RBI$, and wings/core, $WCI$, indicators that quantify the line profile distortions. 

Several parameters are expected to influence the effect of microlensing on those observables. Since the purpose of this study is to discriminate between the different BLR models on the basis of microlensing-induced line profile distortions, it is important to examine the effect on the observables not only of the BLR models but also of other parameters, which most often cannot be fixed or evaluated, such as the caustic pattern, the size of the continuum source, the size of the BLR, the BLR emissivity, and the inclination of the line of sight with respect to the accretion disk + BLR symmetry axis.

\begin{figure}
\centering
\resizebox{\hsize}{!}{\includegraphics*{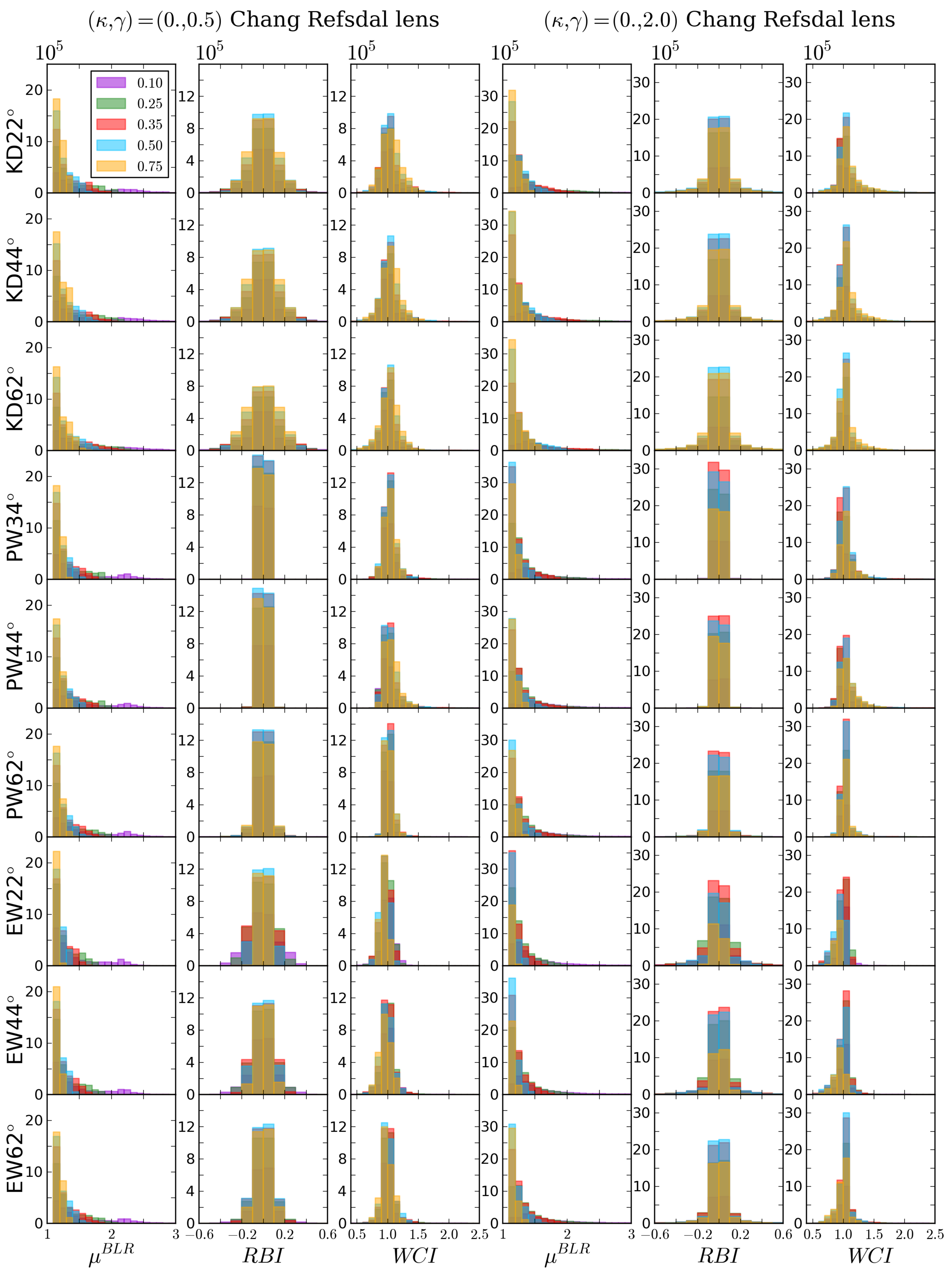}}
\caption{Distributions of BLR magnifications, $\mu^{BLR}$, red/blue indicators, $RBI$, and wings/core indicators, $WCI$, computed for the different BLR models magnified either by a diamond-shaped (Cols. 1 to 3) or by a triangle-shaped Chang-Refsdal caustic (Cols. 4 to 6). The histograms give the number of configurations with different distances and angles of the BLR model relative to the caustic structure that produce a given range of $\mu^{BLR}$, $RBI$, and $WCI$. The abbreviation KD stands for Keplerian disk, PW for polar wind, and EW for equatorial wind. The inclination (in degrees) is denoted next to the initials of the model. The colors correspond to different BLR sizes, $r_{\text{in}}/r_E$. Only configurations with $\mu^{BLR} > 1.1$ are shown.}
\label{fig:distrib_mublr_rbi_wci}
\end{figure}

For each BLR model seen at different inclinations, the distributions of $\mu^{BLR}$, $RBI$, and $WCI$ that result from the magnification either by a diamond-shaped or by a triangle-shaped caustic are illustrated in Fig.~\ref{fig:distrib_mublr_rbi_wci} for several BLR sizes, specifically $r_{\text{in}} = 0.1, \, 0.25, \, 0.35, \, 0.5,$ and $0.75 \, r_E$. A set of representative inclinations is used: i.e., 22\degr, 44\degr\ and 62\degr\ for the Keplerian disk and equatorial wind models, and 34\degr, 44\degr\ and 62\degr\ for the polar wind model (restricted to $i \geq 30\degr$; Sect. \ref{sec:polwind}). Since it is observationally difficult to distinguish the spectral variations due to small-amplitude microlensing effects from those due to noise or intrinsic variations of the quasar propagating in the lensed images with different time delays, our study is restricted to microlensing effects magnifying the line flux by more than $10\%$, i.e., $\mu^{BLR}>1.1$. A first glance reveals that both caustic structures produce $\mu^{BLR}$, $RBI$, and $WCI$ distributions that have comparable shapes. For every BLR model, the $RBI$ distribution is symmetric, which means that red and blue microlensing-induced deformations of the line profiles are equally probable, as expected from the symmetry of the caustic patterns and BLR models. Although sometimes skewed to $WCI > 1$ or $WCI < 1$, the distributions of $WCI$ indicate that microlensing as often affects the wings as the core of the lines.\ The magnification $\mu^{BLR}$ clearly varies with the size of the BLR model, while the geometry and kinematics of the BLR models determine the appearance of the $RBI$ and $WCI$ histograms.

\begin{figure*}
\resizebox{\hsize}{!}{\includegraphics*{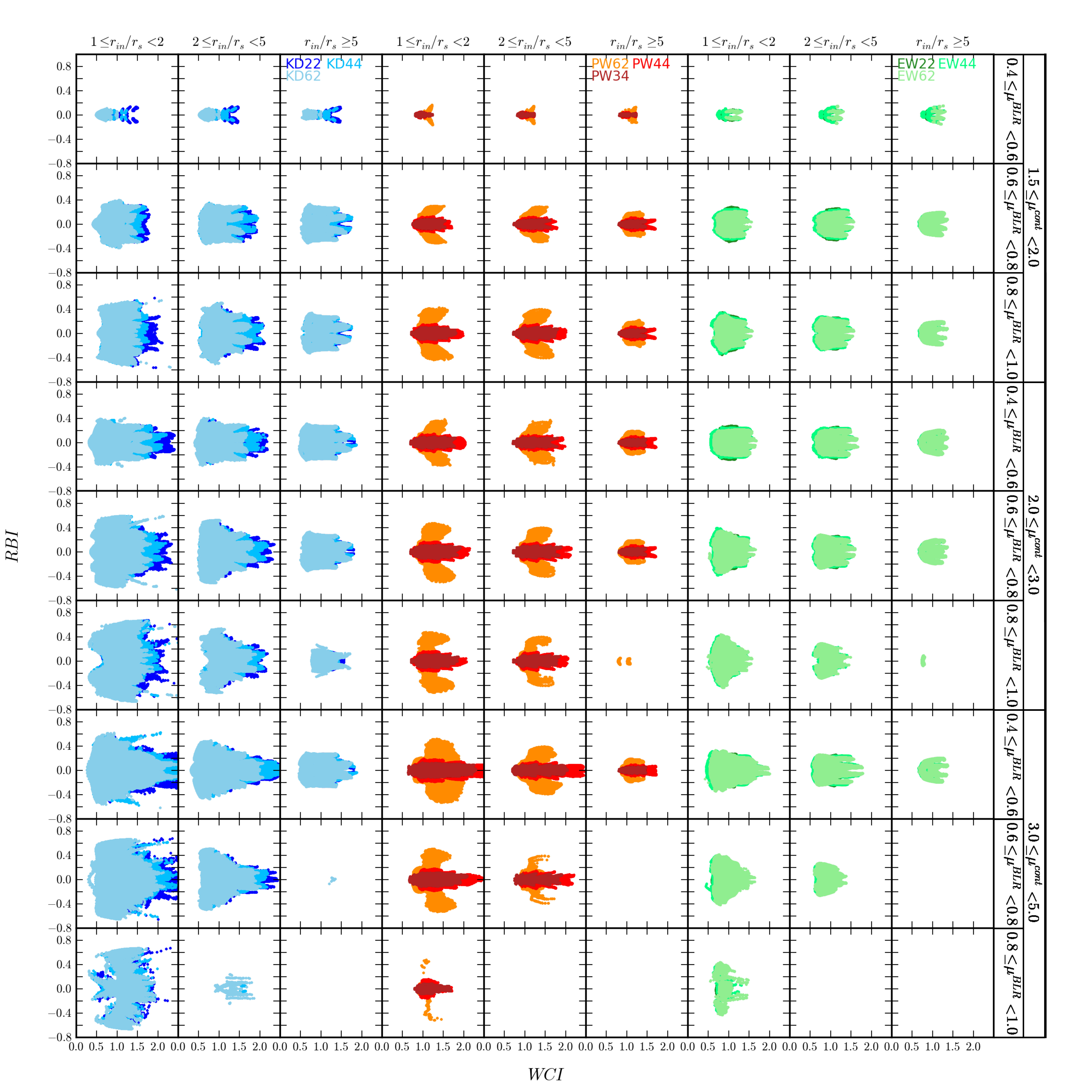}}
\caption{Two-dimensional distributions  $(WCI,RBI)$ that result from microlensing of the BLR models by the diamond-shaped $(\kappa=0, \gamma=0.5)$ Chang-Refsdal caustic. Models in which the emission line emerges at small ($1 \leq r_{\text{in}}/r_s < 2$), large ($2 \leq r_{\text{in}}/r_s < 5$), and much larger ($r_{\text{in}}/r_s \geq 5$) radial distances from the accretion disk are examined separately. The simulations are further divided into several intervals of continuum magnification $\mu^{cont}$ and relative BLR magnification, $\mu^{BLR}$, expressed in units of $\mu^{cont}$; these quantities are specified at the end of each row. The KD model, seen at $22\degr$, $44\degr$, and $62\degr$, is plotted with shades of blue; the EW model, seen at the same viewing angles, is plotted with shades of green; and the PW model, seen at $34\degr$, $44\degr$, and $62\degr$, is plotted in shades of orange and red. The empty squares translate the inability of extended BLR models ($r_{\text{in}}/r_s \geq 5$) to reproduce large-amplitude magnifications of both the continuum and emission line.}
\label{fig:red/blue_effect_g0.5}
\end{figure*}

\begin{figure*}
\resizebox{\hsize}{!}{\includegraphics*{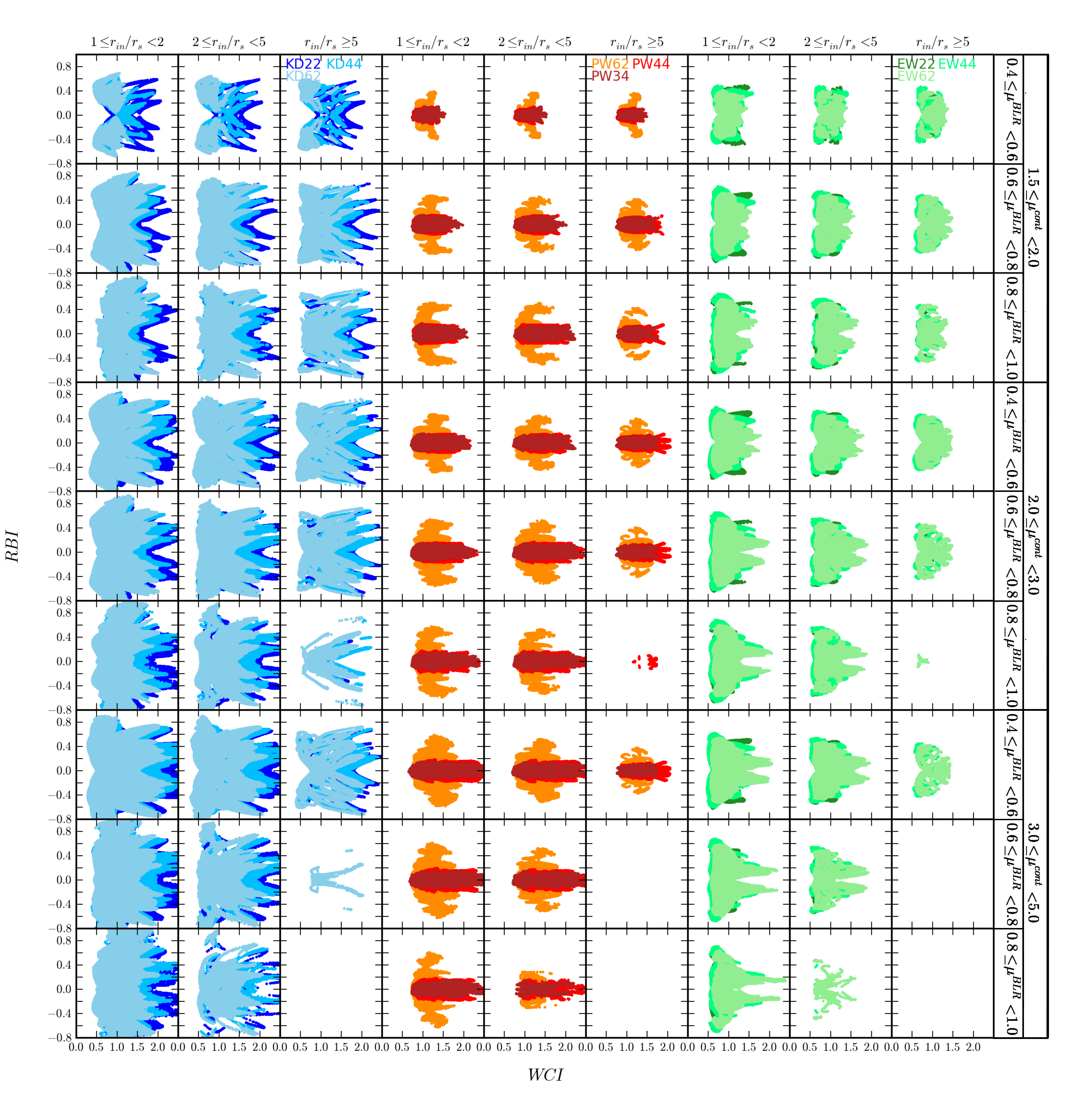}}
\caption{Same as Fig.~\ref{fig:red/blue_effect_g0.5} but for magnification by the triangle-shaped $(\kappa=0, \gamma=2.)$ Chang-Refsdal caustic.}
\label{fig:red/blue_effect_g2.0}
\end{figure*}

A more detailed analysis of the line profile distortions can be achieved by plotting the red/blue indicator as a function of the wings/core indicator. The $(WCI,RBI)$ two-dimensional distributions produced by a diamond-shaped caustic and by a pair of triangle-shaped caustics are illustrated in Figs.~\ref{fig:red/blue_effect_g0.5} and \ref{fig:red/blue_effect_g2.0}, respectively. Three different relative sizes of the BLR with respect to the continuum source are examined, corresponding to emission lines that arise at small ($1 \leq r_{\text{in}}/r_s < 2$), large ($2 \leq r_{\text{in}}/r_s < 5$), and much larger ($r_{\text{in}}/r_s \geq 5$) radial distances from the accretion disk. The $(WCI,RBI)$ distributions are further subdivided as function of the magnification of the continuum emission, $\mu^{cont}$, and of the relative magnification of the emission line, $\mu^{BLR}/\mu^{cont}$. Since no demagnification occurs in the neighborhood of the diamond-shaped caustic, the microlensing configurations that produce a demagnification of the continuum or of the BLR with triangle-shaped caustics are discarded from the analysis to allow a straightforward comparison.

Figures~\ref{fig:red/blue_effect_g0.5} and \ref{fig:red/blue_effect_g2.0} show that different BLR geometries and kinematics lead to different $(WCI,RBI)$ distributions, i.e., to different deformations of the line profiles. This suggests that the measurement of $WCI$ and $RBI$ reported in such ``diagnostic diagrams'' can provide constraints on the BLR models. For example line profile distortions with large  $(WCI,RBI)$ values can only occur in Keplerian disk models. One also immediately notices that large continuum and BLR magnification, i.e., $3 \leq \mu^{cont} < 5$ and $\mu^{BLR}/\mu^{cont} \geq 0.6$, cannot be reproduced by models involving an extended BLR, i.e., $r_{\text{in}}/r_s \geq 5$; the subpanels corresponding to such combinations of criteria thus remain empty. In the following subsections, we discuss the impact of the various parameters in more detail.

\subsection{Impact of the BLR spatial and velocity structure, and its viewing angle}

Figure~\ref{fig:distrib_mublr_rbi_wci} indicates that red/blue asymmetric distortions may be a good discriminant between the polar wind and other models. Indeed, when seen at low inclination, the magnification of a polar wind only produces microlensed line profiles characterized by small red/blue asymmetries, i.e., $\left| RBI \right| <0.1$, whatever the caustic structure. The reason is that the approaching and receding parts of the polar wind partially superimpose in projection and are thus similarly microlensed. This occurs when the line of sight peers into the biconical outflow (PW34$\degr$) or grazes its inner shell (PW44$\degr$). On the contrary, when the line of sight grazes the outer shell of the biconical outflow (PW62$\degr$), asymmetric line profiles can be observed, similar to the Keplerian disk (KD) and equatorial wind (EW) models for which there is a clear-cut separation between the approaching and receding parts of the velocity field (Sect.~\ref{sec:prop_blr})

Large values of $WCI$, which correspond to a magnification of the high-velocity part of the line profile that is higher than the magnification of its low-velocity part, are less likely to be reproduced by models PW62$\degr$ or EW than by models KD or PW at low inclinations. For both caustic structures, configurations resulting in $\mu^{BLR} \geq 1.1$ and $WCI \geq 1.5$ are four times more numerous for the KD, PW34$\degr$, and PW44$\degr$ models than for the PW62$\degr$ and EW models. This is because the high-velocity regions of the PW62$\degr$ and EW models are more extended in projection and therefore less affected by microlensing; this is indicated by the increase of the mean emission radius with the Doppler velocity in EW models (Sect.~\ref{sec:eqwind}) and a mean emission radius that remains larger than $3 \, r_{\text{in}}$ at high Doppler velocities in PW62$\degr$ (Sect.~\ref{sec:polwind}).

The microlensing effect measured through the line velocity structure (i.e., red/blue or wings/core effects) thus clearly depends on the geometry and kinematics of the BLR. However, although they display different shapes, the $(WCI,RBI)$ distributions obtained for the different BLR models superimpose on large parts of the $(WCI,RBI)$ plane, in particular when $WCI \sim 1$ and $RBI \sim 0$. A good discrimination between the BLR models therefore requires sufficiently strong microlensing effects.

The effect of microlensing on the line that arises from the polar wind, and in particular its ability to magnify differentially the blue and red parts of the line profile, also depends on the inclination. On the contrary, the effect of microlensing on the Keplerian disk and equatorial wind models appears insensitive to the inclination. Those effects are better seen in Figs.~\ref{fig:red/blue_effect_g0.5} and \ref{fig:red/blue_effect_g2.0}. The $(WCI,RBI)$ distributions computed for the PW62$\degr$ model are significantly different from those of the PW34$\degr$ and PW44$\degr$ models, whereas they do not change with the inclination for the Keplerian disk and equatorial wind models.

\begin{figure*}
\centering
\resizebox{\hsize}{!}{\includegraphics*{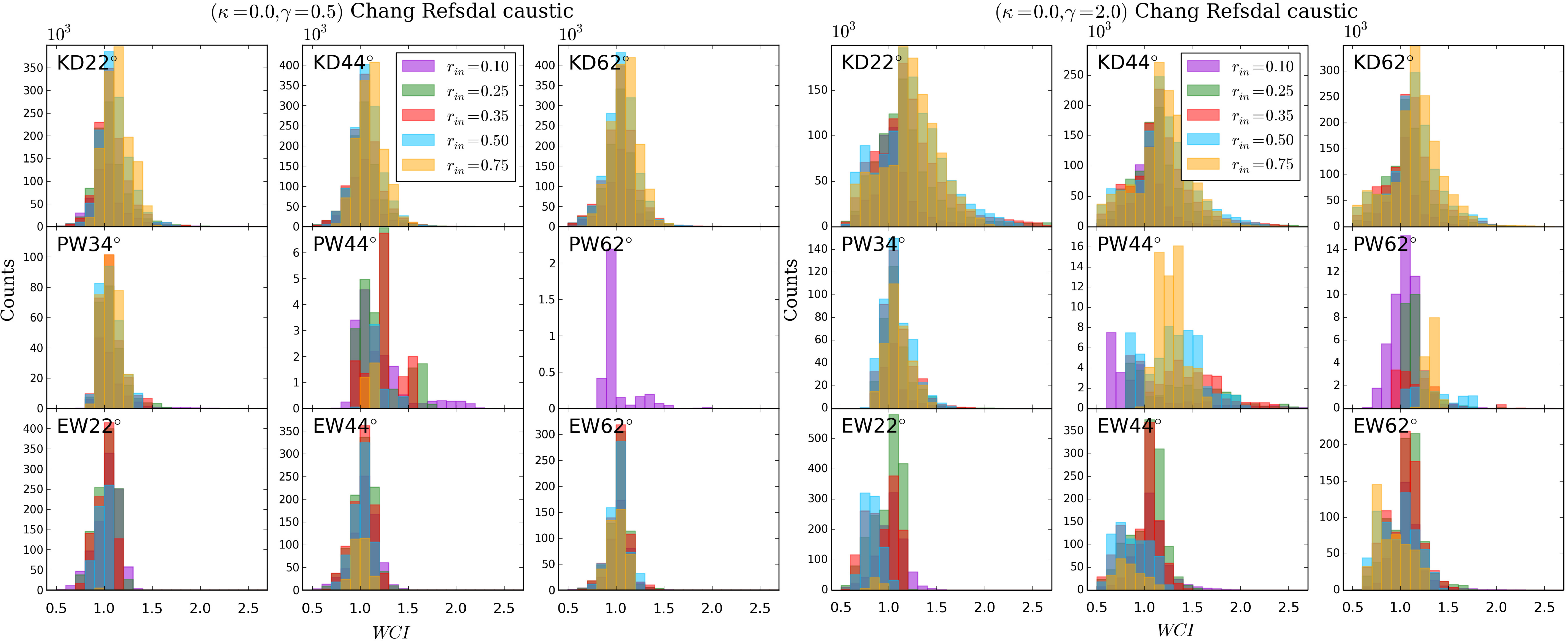}}
\caption{Enlargement of the $WCI$ distribution illustrated in Fig.~\ref{fig:distrib_mublr_rbi_wci}, after selecting asymmetric line profile distortions, i.e., $\left| RBI \right| \geq 0.1$.}
\label{fig:distrib_wci_for_asym}
\end{figure*}

\begin{figure}[t]
\resizebox{\hsize}{!}{\includegraphics*{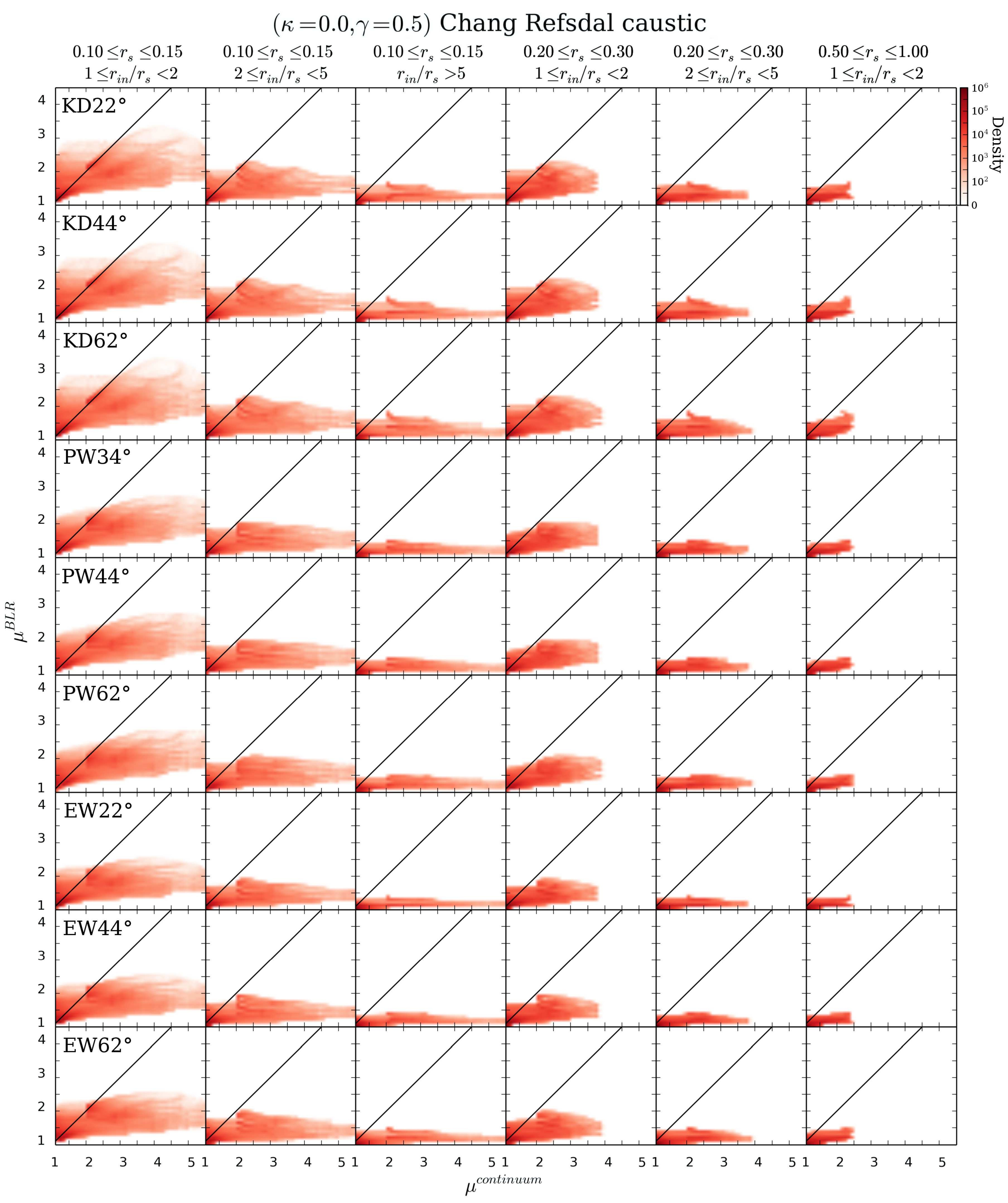}}
\caption{Magnification of the BLR plotted as a function of the magnification of the continuum source produced by a diamond-shaped caustic, for different radii of the continuum source, $r_s$, and different relative sizes of the BLR, $r_{\text{in}}/r_s$. Configurations leading to continuum magnifications $\mu^{cont} < 1.1$ are not considered. }
\label{fig:mublr_mucont_g0.5}
\end{figure}

\begin{figure}[t]
\resizebox{\hsize}{!}{\includegraphics*{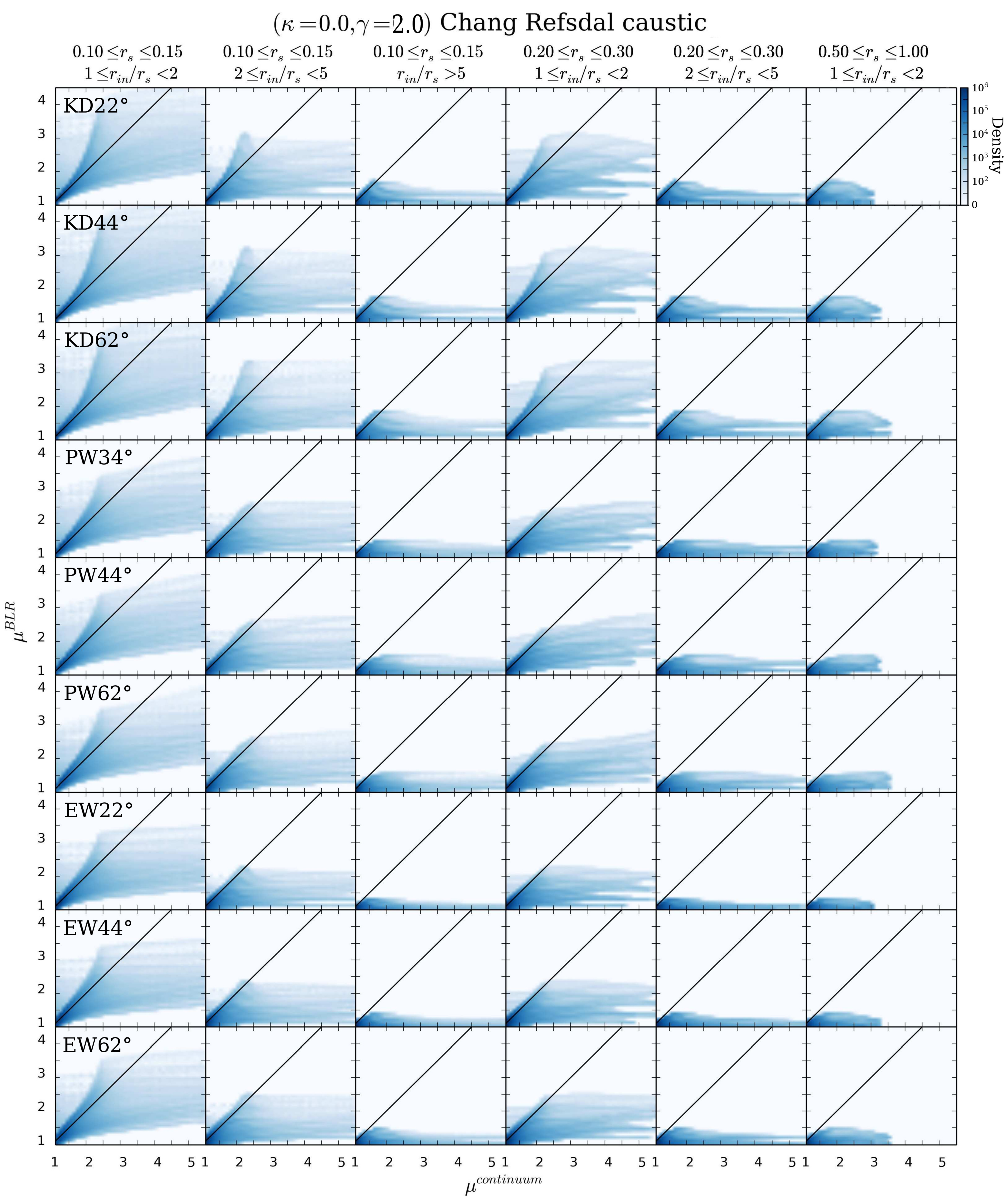}}
\caption{Same as Fig.~\ref{fig:mublr_mucont_g0.5} for the magnification caused by a pair of triangle-shaped caustics. }
\label{fig:mublr_mucont_g2}
\end{figure}

\subsection{Impact of the size of the line and continuum emitting regions}

The distributions of  $\mu^{BLR}$ plotted in Fig.~\ref{fig:distrib_mublr_rbi_wci} show that the amplitude of the magnification decreases for increasing BLR sizes, regardless of the BLR model and caustic structure. Figure~\ref{fig:distrib_mublr_rbi_wci} also shows that, while the distributions of $WCI$ and $RBI$ shrink with increasing BLR size, their shape is not strongly modified. This is in agreement with the behavior of the $(WCI,RBI)$ patterns observed in Figs.~\ref{fig:red/blue_effect_g0.5} and~\ref{fig:red/blue_effect_g2.0} when varying only the relative size of the BLR, $r_{\text{in}}/r_s$. A weak but sensible dependence between the centroid of the $WCI$ distributions and the size of the BLR is nevertheless found when selecting asymmetrical microlensing effects, i.e., $\left| RBI \right| \geq 0.1$, as illustrated in Fig.~\ref{fig:distrib_wci_for_asym}; since PW34$\degr$ and PW44$\degr$ cannot produce as many asymmetric distortions as the other models, they are discarded from the following discussion. We see that the mean $WCI$ increases with the radius of the BLR for the KD and PW62$\degr$ models and  decreases for the EW models. In other words, for increasing BLR sizes, the asymmetric magnification preferentially affects the high-velocity part of the Keplerian disk and polar outflow, and preferentially affects the low-velocity region of the equatorial wind. In the KD and PW62$\degr$ models, the asymmetric magnification of the high-velocity regions is favored by a larger spatial separation between the rapidly approaching and receding parts of the velocity field, which are otherwise close to the center and to each other in projection; this larger spatial separation results from an increase of the BLR inner radius. The EW model shows an opposite behavior because, for increasing BLR sizes, the high-velocity part of the accelerating equatorial outflow, which is located at larger radii, becomes more and more extended, decreasing the magnification of the bluest and reddest parts of the emission line.

\begin{figure*}
\resizebox{\hsize}{!}{\includegraphics*[trim={0 422mm 0 18mm},clip]{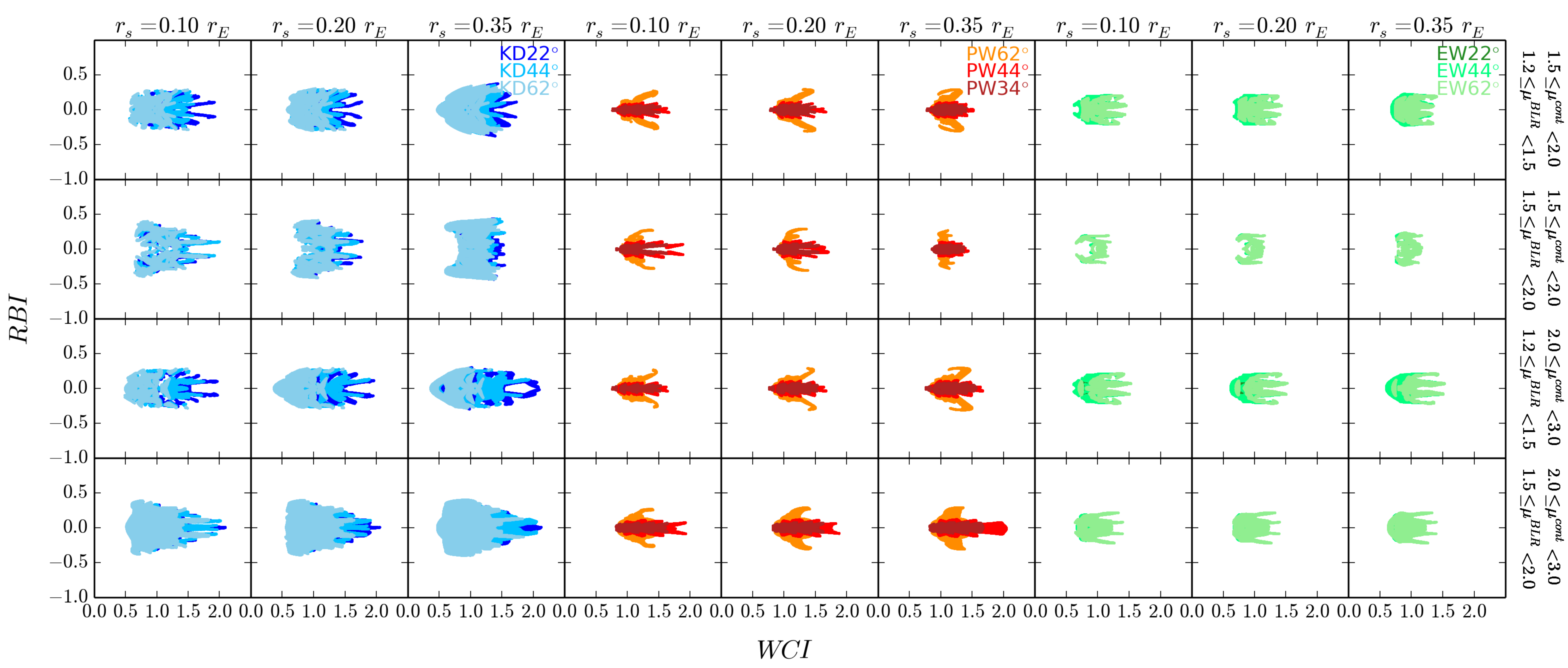}}
\caption{Dependence of the line deformations on the size of the continuum source for a fixed BLR radius  $r_{\text{in}} \, = 0.35 \, r_E$ and the $(\kappa=0,\gamma=0.5)$ Chang-Refsdal caustic. The $(WCI,RBI)$ distribution is illustrated in several intervals of $\mu^{cont}$ and $\mu^{BLR}$.  The radius of the continuum source is varied between $0.1 \, r_E$ and $r_{\text{in}} = 0.35 \, r_E$.}
\label{fig:red/blue-varyingrS_1}
\end{figure*}

\begin{figure*}
\resizebox{\hsize}{!}{\includegraphics*[trim={0 0 0 440mm},clip]{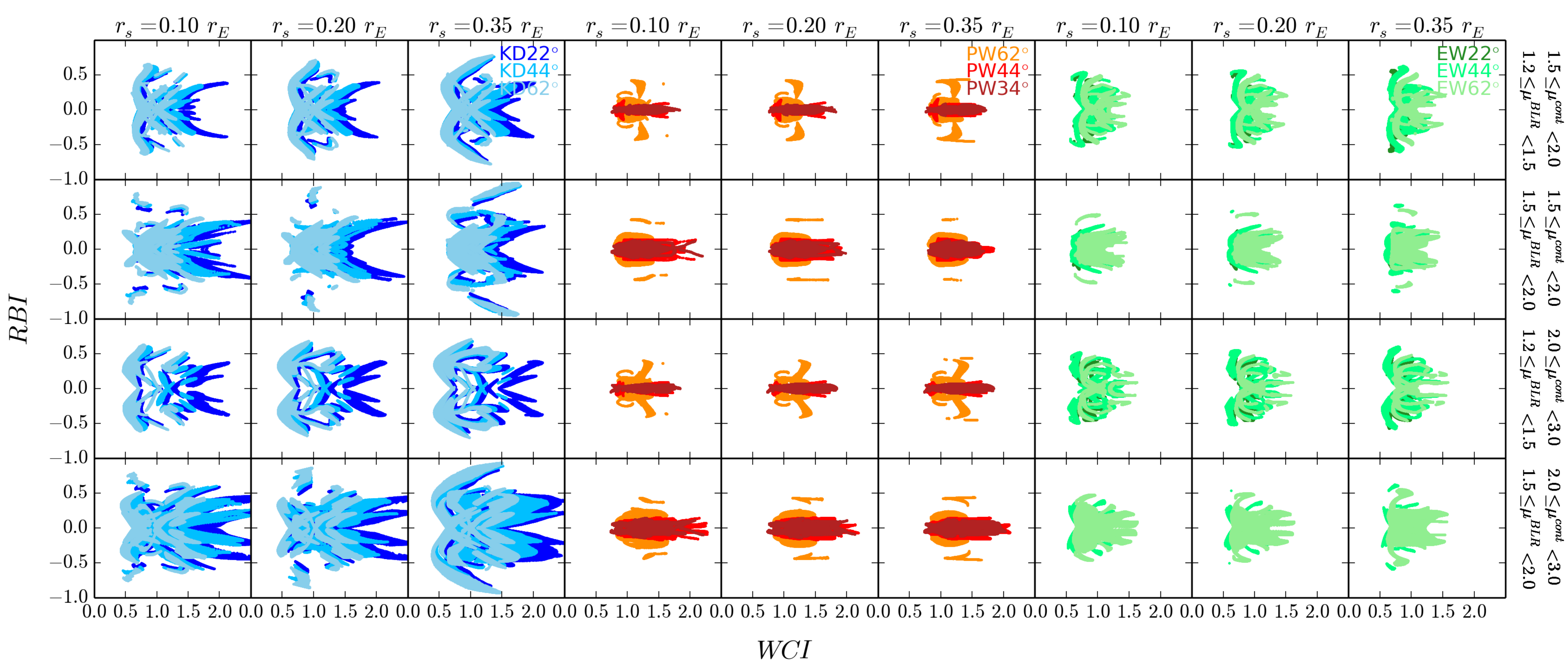}}
\caption{Same as Fig.~\ref{fig:red/blue-varyingrS_1} for the $(\kappa=0,\gamma=2)$ Chang-Refsdal caustic.}
\label{fig:red/blue-varyingrS_2}
\end{figure*}

The radius of the accretion disk is the main parameter controlling the magnification experienced by the continuum emission. The magnification of the BLR is plotted against the magnification of the continuum source in Figs.~\ref{fig:mublr_mucont_g0.5} and \ref{fig:mublr_mucont_g2} for different BLR models, continuum source sizes, $r_s$, and BLR relative sizes, $r_{\text{in}}/r_s$. Despite considerable scatter, there is a correlation between $\mu^{BLR}$ and $\mu^{cont}$ for small BLR sizes, i.e., $r_{\text{in}} < 0.3 \, r_E$, especially when microlensed by a $(\kappa=0,\gamma=2)$ caustic. Larger BLR, i.e., $r_{\text{in}} > 0.5 \, r_E$, display a limited range of BLR magnifications and no apparent correlation with the magnification of the continuum. In fact, a large number of microlensing configurations result in a substantial magnification of the continuum emission while the BLR remains essentially unmagnified. On the other hand, the small BLR models have a non-negligible probability to be more magnified than the continuum  when  $\mu^{cont} \lesssim 1.5$.

The relation between the effect of microlensing on the emission lines and the radius of the accretion disk $r_s$ is explored in Figs.~\ref{fig:red/blue-varyingrS_1} and~\ref{fig:red/blue-varyingrS_2}. For this purpose, the inner radius of the BLR is fixed to $0.35 \, r_E$ and the $(WCI,RBI)$ distributions of the microlensed line profiles are illustrated in intervals of $\mu^{BLR}$ and $\mu^{cont}$, varying only the size of the continuum source. Essentially, the shape of the $(WCI,RBI)$ distributions does not change with the value of $r_s$. On the other hand,  for both caustic patterns and every BLR model, the $(WCI,RBI)$ distribution seems to  shrink when $\mu^{cont}$ decreases at fixed $r_s$. A smaller magnification of the continuum implies either a larger source or a larger distance to the caustics. A larger distance between the continuum source and the caustics causes the caustics to sample outer, less luminous, regions of the BLR, then resulting in smaller $WCI$ and $RBI$.

\begin{figure*}
\resizebox{\hsize}{!}{\includegraphics*{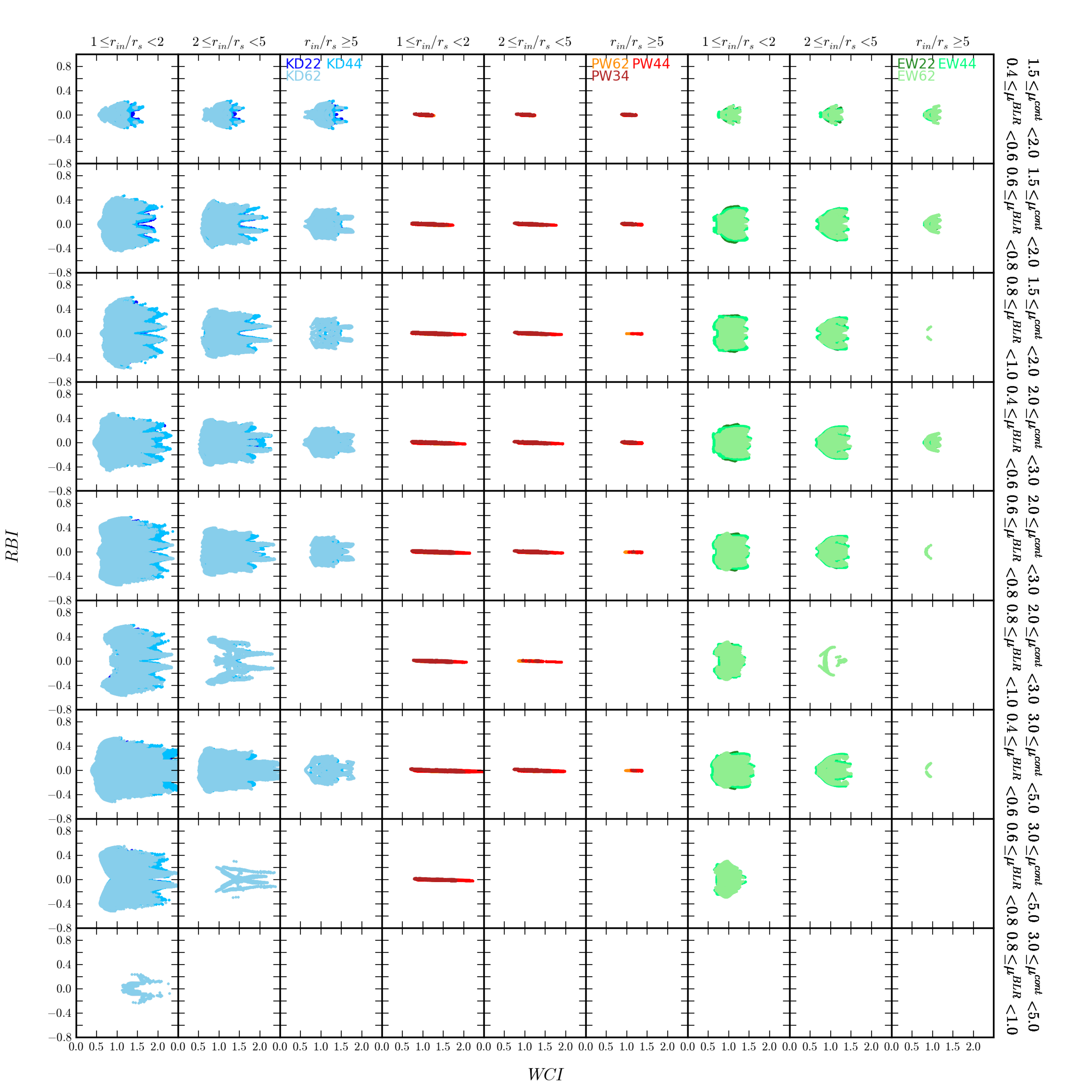}}
\caption{Same as Fig.~\ref{fig:red/blue_effect_g0.5} but with the emissivity law $\epsilon_0 \, (r/r_{\text{in}})^{-1.5}$.}
\label{fig:red/blue_effect_g0.5_q1.5}
\end{figure*}

\begin{figure*}
\resizebox{\hsize}{!}{\includegraphics*{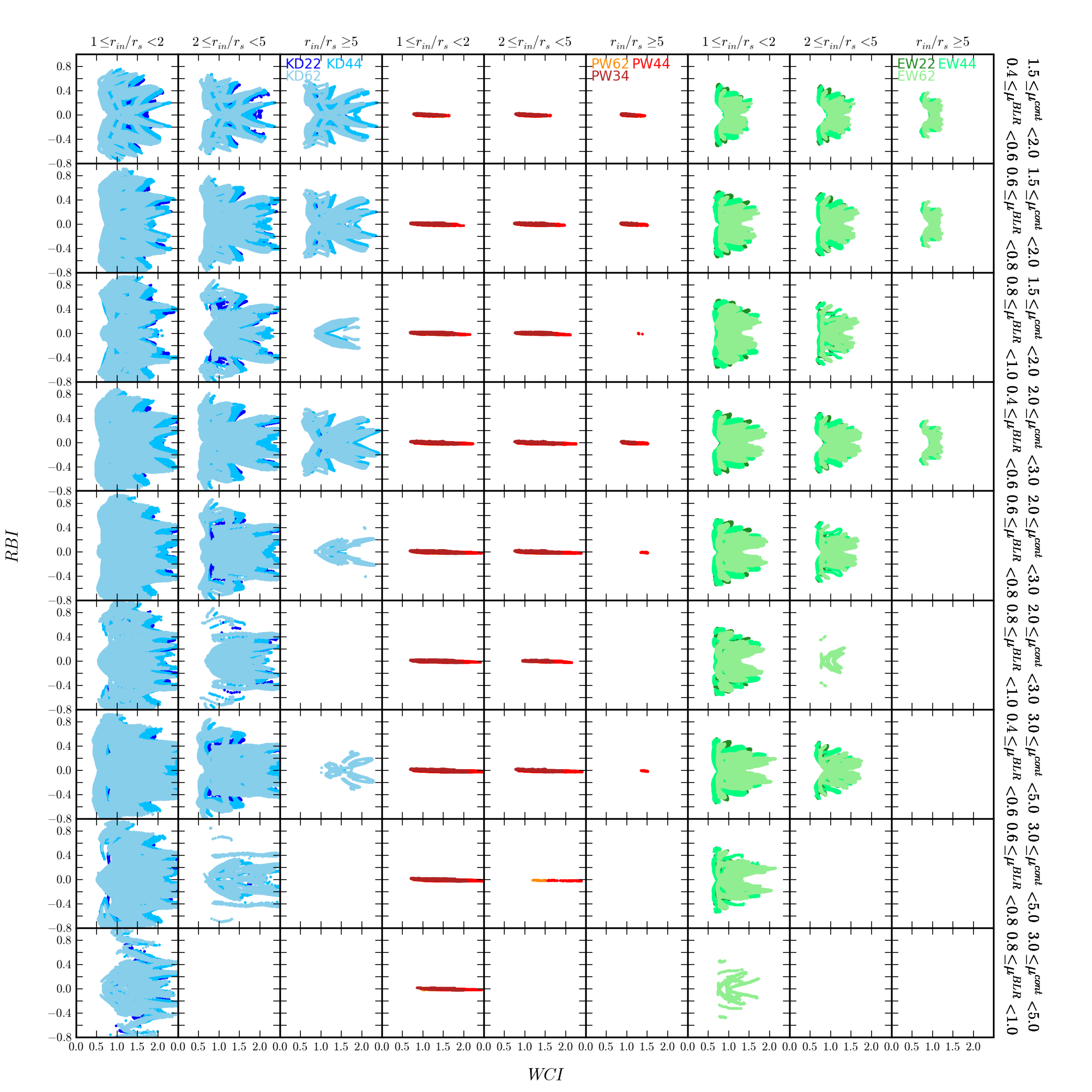}}
\caption{Same as Fig.~\ref{fig:red/blue_effect_g2.0} but with the emissivity law $\epsilon_0 \, (r/r_{\text{in}})^{-1.5}$.}
\label{fig:red/blue_effect_g2.0_q1.5}
\end{figure*}

\subsection{Impact of the caustic structure}

Both the diamond-shaped and triangle-shaped caustics produce $RBI$ and $WCI$ distributions with similar extents, although the frequency of intermediate $RBI$ and $WCI$ values differs, in particular for the Keplerian disk models (Fig.~\ref{fig:distrib_mublr_rbi_wci}).

Noticeable differences appear  when plotting $RBI$ as a function of $WCI$ (Figs.~\ref{fig:red/blue_effect_g0.5} and \ref{fig:red/blue_effect_g2.0}). While both types of caustics are efficient at producing significant asymmetry in the line profile, i.e., $\left| RBI \right| \geq 0.1$, the $(WCI,RBI)$ two-dimensional distributions also show that the triangle-shaped caustic can induce line profile distortions characterized by larger $RBI$, and thus more asymmetric, than the diamond-shaped caustic, especially for the Keplerian disk and the equatorial wind models. The compact triangle-shaped caustic indeed samples a smaller region in the source plane, which can exacerbate the contrast between the magnification of the negative and positive parts of the velocity field.

\subsection{Impact of the emissivity law}
\label{sec:impact_q}

In the previous sections, we considered models in which the emissivity rapidly decreases with the radial distance (Sect.~\ref{sec:emissivity}). One could then  wonder about the impact of a slower varying emissivity on the previous results.

To estimate the impact of the emissivity law, we investigated identical BLR models but we modify the power-law exponent to $q=1.5$ instead of $q=3$. In that case, there is significant emission arising at the outer radius of the BLR model, i.e., $r_{\text{out}} = 10 \, r_{\text{in}}$ (Fig.~\ref{fig:velo+emissivity}). Such a change of emissivity has virtually no impact on the $(WCI,RBI)$ distributions computed for the Keplerian disk, and only results in a slight shrinking of the distributions computed for the equatorial wind. However, significant asymmetries of the microlensed line profile ($\left|RBI \right| > 0.1$) do not occur anymore for the polar wind, regardless of the magnification pattern and the inclination at which the polar wind is seen (Figs.~\ref{fig:red/blue_effect_g0.5_q1.5} and \ref{fig:red/blue_effect_g2.0_q1.5}). This contrasts with the results obtained with $q=3$, where asymmetric deformations were observed for the polar wind seen at $i=62\degr$ (Figs.~\ref{fig:red/blue_effect_g0.5} and \ref{fig:red/blue_effect_g2.0}).

\section{Conclusions}

The effect of gravitational microlensing on quasar broad emission line profiles and their underlying continuum has been simulated considering simple representative BLR models and caustic patterns. Keplerian disks along with polar and equatorial wind BLR models of various sizes have been considered.  The effect of microlensing has been quantified using four observables: $\mu^{BLR}$, the total magnification of the broad emission line; $\mu^{cont}$, the magnification of the underlying continuum; as well as red/blue, $RBI$, and wings/core, $WCI$, indices that characterize the line profile distortions. Those observables were designed to not depend on the exact profile of the BELs, so that they can be directly compared to observations.

The simulations show that asymmetric distortions of the broad line profiles such as those observed in several lensed quasars \citep[e.g.,][]{2012Sluse,2014Braibant,2016Braibant} can indeed be reproduced and attributed to the differential effect of microlensing on spatially and kinematically separated regions of the BLR.

Since the largest differences between microlensing-induced distortions of the line profiles are seen between BLR models of different geometry and kinematics, microlensing measurements can thus help to constrain the BLR structure. In particular, red/blue asymmetric distortions constitute a good discriminant between the polar wind and other models since red/blue asymmetries are never observed when the polar outflow is seen at low inclination, regardless of the BLR size, its emissivity, and the caustic structure. The polar wind is the only BLR model in which the deformations of the line profiles, in particular the differential magnification of the red and blue wings of the line, can depend on the inclination with respect to the line of sight. Asymmetric effects are completely inhibited when the emissivity of the polar wind decreases slowly with the radial distance.

While the spatial and velocity structure of the BLR mostly determines the effect of microlensing through the line profile, the amplitude of the magnification essentially depends on the size of the BLR. The more extended the emission region, the more limited the range of possible magnification  effects. The BLR is indeed magnified by the large-scale caustic structure, which is constituted by areas of low magnification separated by narrow, highly magnifying regions, so that large BLR smear out the microlensing effects. However, the more compact continuum source samples the small-scale caustic structure of the magnification maps. The magnification of the continuum emission can thus be much larger. 

The caustic pattern has a limited influence on the results. Indeed, the profile deformations produced by the diamond-shaped and triangle-shaped Chang-Refsdal caustics qualitatively agree, in the sense that if the diamond-shaped caustic can differentially magnify the line red and blue wings for a given BLR model, a similar effect is caused by the triangle-shaped caustic. However, the compact triangle-shaped caustic appears to exacerbate the amplitude of differential red/blue effects compared to the diamond-shaped caustic. The triangle-shaped caustic indeed samples a smaller region of the source plane, which makes it more likely to selectively magnify the approaching or receding parts of the BLR, especially in the Keplerian disk and equatorial wind models that show a clear-cut separation between the positive and negative parts of the velocity field.

The magnification of the emission line $\mu^{BLR}$ sets an upper limit on the BLR size and, similarly, the magnification of the continuum $\mu^{cont}$  sets an upper limit on the size of the continuum source. Since the line profile distortions mainly depend on the BLR geometry and kinematics, the measurement of the microlensing effect through the velocity structure of the line is the most promising way to discriminate between BLR models.  In particular, the $(WCI,RBI)$ diagrams could serve as ``diagnostic diagrams'' to disentangle the BLR models on the basis of quantitative measurements. However, there is a large overlap between the microlensing-induced line distortions produced for the different BLR models, particularly when the effect has a weak amplitude.  We then expect that a good discrimination between the BLR models would require sufficiently strong microlensing effects, at least when only based on single epoch measurements. To better reproduce the observations, more complex models of the BLR that combine the simple geometries and kinematics investigated here might be necessary and,  in particular, velocity fields that change from rotation-dominated to wind-dominated with the distance to the core. It is likely that only a long-term and high-frequency spectrophotometric monitoring of suitable targets will allow us to disentangle more complex models.

In a future work, we will confront the results of these simulations, in particular the $(WCI,RBI)$ diagrams, to the line profile distortions actually observed in the lensed quasars HE0435-1223 and Q2237+0305 \citep{2014Braibant, 2016Braibant}.

\bibliographystyle{aa}
\bibliography{references}

\end{document}